\documentclass[prc,twocolumn,eprint=false]{revtex4-1}

\usepackage{graphicx}%
\usepackage{dcolumn}%
\usepackage{bm}
\usepackage{slashed}
\usepackage{amsmath,graphicx}
\usepackage[colorlinks=true,linktocpage=true,linkcolor=blue,citecolor=blue]{hyperref}
\usepackage{float}
\usepackage{nicefrac}
\usepackage[normalem]{ulem}
\usepackage{amsmath}
\usepackage[caption=false]{subfig}

\def\k{{\boldsymbol k}}

\def\dd{{\rm d}}

\def\ku{{k\cdot u}}
\def\trel{\tau_{\rm rel}}

\def\vsound{c_s}
\def\DK{{\rm d} K}

\def\Eps{\varepsilon}
\def\PL{{P_\parallel}}
\def\PT{{P_\perp}}
\def\P{P}
\def\eff{{\tiny{\rm eff}}}

\def\fenez{f_\varepsilon(z)}
\def\fpez{f_P(z)}
\def\aepsn{a^{(n)}_\varepsilon}
\def\apen{a^{(n)}_P}
\def\aetan{a^{(n)}_\eta}
\def\azeta1{a^{(n)}_{\zeta_1}}
\def\azetad{a^{(n)}_{\zeta_2}}

\def\s2{\sqrt{2}\,}
\def\Lisn12{{\rm Li}_{n+\frac{1}{2}}\left(e^{-\s2 a} \right) }
\def\Lisnm12{{\rm Li}_{n-\frac{1}{2}}\left(e^{-\s2 a} \right) }

\newcommand{\onethird}{{\nicefrac{1}{3}}}

\newcommand{\smallG}{{\rm \scriptscriptstyle{G}}}

\newcommand{\GZ}{{\rm \scriptscriptstyle{GZ}}}
\newcommand{\SB}{{\rm \scriptscriptstyle{SB}}}
\newcommand{\BE}{{\rm \scriptscriptstyle{BE}}}
\newcommand{\beq}{\begin{eqnarray}}
\newcommand{\eeq}{\end{eqnarray}}
\newcommand{\be}{\begin{eqnarray*}}
\newcommand{\ee}{\end{eqnarray*}}
\newcommand{\bqa}{\begin{eqnarray}}
\newcommand{\eqa}{\end{eqnarray}}

\begin{document}

\title{Transport coefficients of the Gribov-Zwanziger plasma}

\author{Wojciech Florkowski}
\email{wojciech.florkowski@ifj.edu.pl}

\author{Radoslaw Ryblewski}
\email{radoslaw.ryblewski@ifj.edu.pl}

\affiliation{The H. Niewodnicza\'nski Institute of Nuclear Physics, Polish Academy of Sciences, PL-31342 Krak\'ow, Poland}

\author{Nan Su}
\email{nansu@th.physik.uni-frankfurt.de}
\affiliation{Institut f\"ur Theoretische Physik, Goethe-Universit\"at Frankfurt, D-60438 Frankfurt am Main, Germany}

\author{Konrad Tywoniuk}
\email{konrad.tywoniuk@cern.ch}
\altaffiliation[Presently at: ]{Theoretical Physics Department, CERN, Geneva, Switzerland}
\affiliation{Deptartamento d'Estructura i Constituents de la Mat\`eria, Institut de Ci\`encies del Cosmos (ICCUB), Universitat de Barcelona, Mart\`i Franqu\`es 1, E-08028 Barcelona, Spain}

\begin{abstract}
We study dynamic features of a plasma consisting of gluons whose infrared dynamics is improved by the Gribov-Zwanziger quantization. This approach embodies essential features of color confinement which set the plasma apart from conventional quasiparticle systems in several aspects. Our study focusses on a boost-invariant expansion for in- and out-of-equilibrium settings within the relaxation time approximation, which at late times can be characterized by the sound velocity, $c_s$, and the shear, $\eta$, and bulk, $\zeta$, viscosities. We obtain explicit expressions for the transport coefficients $\eta$ and $\zeta$ and check that they are consistent with the numerical solutions of the kinetic equation. At high temperature, keeping both the Gribov parameter and the relaxation time constant, we find a scaling $\zeta/\eta \propto \onethird - c_s^2$ which manifests strong breaking of conformal symmetry in contrast to the case of weakly coupled plasmas.
\end{abstract}


\maketitle

\section{Introduction}

The fate of hadronic matter at high temperature is still under active investigation nearly five decades after the theory of quantum chromodynamics (QCD) was established. Above a certain (pseudo-)critical temperature, the fundamental quark and gluon constituents are expected to become the active degrees of freedom, forming a quark-gluon plasma (QGP). At asymptotically high energies, such a plasma has been for long regarded as a weakly interacting gas. However, this naive picture has been challenged by experimental data \cite{ALICE:2011ab,Aamodt:2011by,Chatrchyan:2012wg,Aad:2013xma,Adamczyk:2013waa,Adare:2011tg} from ultrarelativistic heavy-ion collisions that give evidence for the creation of a strongly  interacting and correlated system in the accessible energy regime \cite{Gyulassy:2004zy,Shuryak:2004cy}. This has raised the interest in describing the dynamics of the collisions in terms of dissipative fluid dynamics, characterized by various transport coefficients \cite{Romatschke:2009kr,Jeon:2015dfa}, see also \cite{Gale:2013da}. These, in turn, are sensistive to the long-wavelength features of the underlying microscopic theory by force of the Green-Kubo relations \cite{Jeon:1995zm}.

While the thermodynamic properties of QCD are well established at high temperature, both within resummed perturbation theory, see \cite{Andersen:2004fp,Su:2012iy,Su:2015esa} for reviews and \cite{Andersen:2010ct,Andersen:2011sf,Mogliacci:2013mca,Haque:2014rua} for the most up-to-date results on thermodynamic quantities, and by lattice methods \cite{Bazavov:2014pvz,Borsanyi:2013bia}, understanding the dynamic properties remains challenging \cite{Meyer:2011gj}. Let us point out the two main reasons. In the small-coupling regime, $g\ll 1$, the  degrees of freedom successfully described by resummed perturbation theory are dressed quasiparticles with an effective mass $\sim g T$ \cite{Blaizot:2001nr,Arnold:2002zm,Hong:2010at,Moore:2008ws},~where $g$ is the QCD coupling constant. 
The perturbative QCD predictions are however in tension with experimental results since the expected running-coupling strength in the phenomenologically relevant regime is of the order of unity, $g \sim \mathcal{O}(1)$ \cite{Andersen:2004fp,Su:2012iy,Su:2015esa}.
Secondly, since the initial conditions of ultrarelativistic heavy-ion collisions are highly anisotropic in momentum space, the incipient evolution of the created system towards a state complying with a hydrodynamic description is still debated and new formulations of hydrodynamics are constructed in this context \cite{Florkowski:2010cf,Martinez:2010sc,Strickland:2014pga}. These outstanding issues have in turn motivated the application of strong-coupling techniques, based on the gauge-gravity duality, see e.g. \cite{CasalderreySolana:2011us}. One of the striking hallmarks of the latter approaches is the lack of distinct quasiparticle excitations in the spectrum of the theory, but rather the dominance of hydrodynamic sound modes \cite{Kovtun:2006pf,Teaney:2006nc,Heller:2014wfa}.

In our present work, we will focus on two main transport coefficients for soft modes in a charge-free plasma, namely the shear, $\eta$,  and bulk, $\zeta$, viscosities.  In the last years, substantial knowledge has been gained regarding their properties. 
In particular, the shear and bulk viscosities were calculated in the high-$T$ regime using perturbation theory in \cite{Arnold:2000dr,Arnold:2003zc} and \cite{Arnold:2006fz,Moore:2008ws}, respectively. 
In the phenomenologically relevant temperature regime, $T \sim (1-4)T_c$ \cite{Su:2012iy,Su:2015esa}, lattice calculations are still plagued by large uncertainties, see \cite{Meyer:2007dy} for early results on bulk and \cite{Meyer:2007ic} for shear viscosities, respectively. 
One noteworthy feature is the enhancement of the bulk viscosity close to the critical temperature, which was also found in~\cite{Kharzeev:2007wb,Karsch:2007jc}. 
Encouragingly, the transport coefficients of the Yang-Mills theory have been recently computed using functional renormalization techniques \cite{Haas:2013hpa,Christiansen:2014ypa}. 
In a complementary effort, the transport properties of supersymmetric plasmas at high temperatures have recently been studied using the gauge-gravity duality. For a conformal system a lower bound on the shear viscosity to entropy  density ratio, $\eta/s \geq 1/4\pi$, was found \cite{Policastro:2001yc}. However, breaking of conformal invariance is necessary for the generation of bulk viscosity \cite{Benincasa:2005iv,Buchel:2005cv}, and several models have been proposed in order to mimic lattice QCD features \cite{Finazzo:2014cna,Li:2014dsa}.

Due to confinement effects, the infrared (IR) regime of QCD is strongly coupled, opposite to the case of quantum electrodynamics.
Long-range correlations in the system contribute to the confinement of colored degrees of freedom.
These features are genuinely non-perturbative and are therefore beyond the scope of conventional perturbative techniques \cite{Andersen:2004fp,Su:2012iy,Su:2015esa}, especially in the study of the QGP~\cite{Linde:1980ts,Gross:1980br} and for heavy-ion phenomenology.

Motivated by these considerations, in this work we continue our investigation of equilibrium and non-equilibrium properties of a plasma consisting of gluons obtained from the Gribov-Zwanziger (GZ) quantization which was introduced in Gribov's seminal paper \cite{Gribov:1977wm} and later systemized by Zwanziger in~\cite{Zwanziger:1989mf}, for reviews see \cite{Dokshitzer:2004ie,Vandersickel:2012tz}. 
This prescription improves the IR behavior of Yang-Mills theory by fixing residual gauge transformations that remain after applying the perturbative Faddeev-Popov procedure. 
Consequently, a new scale $\gamma_\smallG$, the Gribov parameter, is introduced, which leads to an IR-improved dispersion relation for gluons \cite{Gribov:1977wm}.  In the Coulomb gauge this dispersion relation reads
\beq
\label{eq:GZdispersionNonCov}
E(\k) = \sqrt{\k^2 + \frac{\gamma_\smallG^4}{\k^2}} \,,
\eeq
where $\k$ is the three-momentum and $E$ is the energy. It embodies the expected behavior: a large energy cost associated with soft gluons, which subsequently amounts to the reduction of the physical state space \cite{Gribov:1977wm,Zwanziger:1989mf}, see also \cite{Feynman:1981ss}. The Gribov parameter is an intrinsic Yang-Mills scale that is bootstrapped by a self-consistent gap equation \cite{Gribov:1977wm,Zwanziger:1989mf} and explicitly breaks the conformal symmetry of the theory.

This framework has recently attracted a lot of attention in the theory community. In particular, an excellent agreement between Gribov's result and lattice data in the Coulomb gauge has been established for the gluon propagator \cite{Burgio:2008jr}. As one of the most important features, it was early realized that the positivity of the gluon spectral function is violated in this theory at any temperature, see \cite{Alkofer:2000wg,Maas:2011se} for reviews. This implies the absence of confined particles in the physical asymptotic spectrum and is thus a crucial measure of color confinement. 

The Gribov approach was generalized to finite temperature \cite{Zwanziger:2004np} and it was shown that the Gribov parameter provides a natural QCD scale  that significantly improves the IR behavior of the theory and provides good agreement with lattice results on thermodynamic quantities~\cite{Fukushima:2013xsa}. Owing to its intimate relation to the (chromo)magnetic scale \cite{Zwanziger:2006sc,Fukushima:2013xsa}, collective massless degrees of freedom are present at finite temperature \cite{Su:2014rma}, see also \cite{Chernodub:2007rn,Cooper:2015bia}. Recently, the connection between the Gribov parameter and the topological structure of the QCD vacuum has been explored, shedding light on a profound relationship between the GZ quantization and the confinement/deconfinement phase transition \cite{Kharzeev:2015xsa}, see also \cite{Canfora:2015yia}. The impact on real-time observables has also been examined lately \cite{Bandyopadhyay:2015wua}.

Motivated by the results discussed above, the Gribov dispersion relation thus provides a unique and straightforward way to study the impact of residual confinement effects on QGP transport properties. It was for the first time employed in the kinetic and hydrodynamic calculations in our previous work \cite{Florkowski:2015rua}, where the analyzed system was dubbed the Gribov-Zwanziger (GZ) plasma. In addition to obtaining a qualitatively good agreement with the Yang-Mills lattice data \cite{Borsanyi:2012ve} in equilibrium, we calculated the bulk viscosity in a boost-invariant setup \cite{Bjorken:1982qr} using the relaxation-time approximation. The latter quantity attracts growing attention in the literature due to possible importance of bulk viscous effects in relativistic heavy-ion collisions \cite{Fries:2008ts,Bozek:2009dw,Monnai:2009ad,Denicol:2009am,Noronha-Hostler:2013gga,Ryu:2015vwa,Torrieri:2008ip,Rajagopal:2009yw,Floerchinger:2015efa}.

In this paper, we give a detailed discussion of the results obtained in~\cite{Florkowski:2015rua} and, in addition, we present several new and complementary results.  We study the approach to equilibrium of the GZ plasma by varying the characteristic relaxation time, $\trel$, and demonstrate that the late-time behavior of the system can be quantified using transport coefficients. In particular, we calculate the bulk and shear viscosities of the GZ plasma and derive their low- and high-$T$ limits. 

One of our important results is the linear scaling relation $\zeta / \eta \simeq \kappa_\GZ (\onethird - c_s^2)$ for the GZ plasma. This relation holds for constant Gribov parameter and constant relaxation time, and is characterized by the universal values $\kappa_\GZ = 5$ when $T\to 0$ and $\kappa_\GZ = 5/2$ for $T\to \infty$, that are independent of $\gamma_\smallG$. In the intermediate, phenomenologically relevant temperature range the scaling can be approximated by a slowly varying coefficient $\kappa_\GZ$. A similar scaling law was observed from the gauge-gravity duality \cite{Benincasa:2005iv,Buchel:2005cv}, and differs from the expected high-$T$ scaling obtained within perturbative QCD \cite{Arnold:2006fz}.

Our results are compared with the formulas characterizing a plasma of massive particles obeying Bose-Einstein (BE) statistics. 
It turns out that certain physical observables (such as, for example, the speed of sound) have an apparently similar
behavior for the GZ plasma and the BE massive plasma, provided the value $\sqrt{2} \gamma_\smallG$ is adopted for the particle mass. 
However, simple relations do not hold in most cases. The observed differences reflect the modified IR dynamics that sets the GZ plasma apart from conventional quasiparticle models.

The structure of the paper is as follows. In Sec.~\ref{sect:impl} we discuss the implementation of Lorentz covariance and boost-invariance in our model. The implementation of covariance into the model is a fundamental problem connected with the use of the non-covariant dispersion relation, derived by Gribov in the Coulomb gauge, as a starting point. On the other hand, the implementation of boost invariance \cite{Bjorken:1982qr}  is a simplifying assumption that facilitates the analysis of a non-equilibrium system. In Sec.~\ref{sect:firstord}, using the concepts of the underlying kinetic theory, we discuss the form of the hydrodynamic equations for boost-invariant and transversally homogeneous systems, denoted below as (0+1)D systems. In Sec.~\ref{sec:KinEq} we present the kinetic equation in the relaxation time approximation and show its predictions for various physical observables. The forms of the bulk and shear viscosity coefficients are derived in Sec.~\ref{sect:kinco}, where we discuss also the ratio of the two viscosities and relation of $\zeta/\eta$ to the speed of sound. In Sec.~\ref{sec:momdep} we also comment on the effect of the momentum-dependent relaxation time on our results. We summarize and conclude in Sec.~\ref{sec:Conclusions}. The paper is closed with a series of appendices discussing the implementation of Lorentz covariance, various auxiliary functions and the low- and high-temperature expansions for thermodynamic quantities, the speed of sound, shear and bulk viscosities and their ratio.

Throughout the paper we use the natural system of units with $c=\hbar=k_B=1$. The three-vectors are denoted by the bold font, the four-vectors are in standard font, the dot denotes the scalar product with the metric \mbox{$g^{\mu\nu}={\rm diag}(+1,-1,-1,-1)$}. The four-vector defining the heat-bath, or the fluid element's local rest frame (LRF), is denoted by $u$. In this frame $u^{\mu}=(1,0,0,0)$.

\section{Implementation of covariance and boost-invariance}
\label{sect:impl}

\subsection{Dispersion relation in covariant form}

The Gribov dispersion relation, Eq.~(\ref{eq:GZdispersionNonCov}), may be cast into the following covariant form~\cite{Florkowski:2015rua}
\begin{eqnarray}
\label{eq:GZdispersionCov}
E(k \cdot u) = \sqrt{ (k \cdot u)^2 + \frac{\gamma_\smallG^4}{(k \cdot u)^2} } \,,
\end{eqnarray}
where $u$ is the four-velocity of the fluid element. We assume that the Coulomb gauge as well as the in-medium value of $\gamma_\smallG$ are fixed in LRF. We introduce $k^0 \equiv |\k|$, which is the magnitude of the three-vector $\k \equiv (k_x, k_y, k_\parallel)$, and $k_\perp = \sqrt{k_x^2+k_y^2}$ such that the resulting four-vector $k^{\mu} = (k^0, \k)$ has standard Lorentz transformation properties with $ k^2 = 0$. 

It is important to note that the original dispersion relation in Eq.~(\ref{eq:GZdispersionNonCov}) is derived in the Coulomb gauge which explicitly breaks Lorentz invariance. In order to regain a covariant formalism, which is necessary for description of relativistic fluids, one has to make certain assumptions regarding the Lorentz transformation properties of the quantities appearing in Eq.~(\ref{eq:GZdispersionNonCov}). 
This problem is presented and discussed in more detail in Appendix~\ref{sect:cov}.  Our choice, based on Eq.~(\ref{eq:GZdispersionCov}), is the most natural
and free from mathematical ambiguities. We note that in this case  the zeroth component of the momentum, $k^0$, is different from $E(k)$ appearing in the Gribov-Zwanziger formalism \cite{Gribov:1977wm}. The latter is the energy of the interacting particle with a three-momentum~$\k$ in the reference frame of the heat bath.

\subsection{Thermodynamic-like quantities}

In LRF, the energy density of the GZ plasma described by the non-equilibrium momentum distribution function $f(\k)$ is described by the formula~\cite{Zwanziger:2004np}
\beq
\label{eq:EpsNonCov}
\varepsilon = g_0\int \frac{\dd^3k}{(2\pi)^3} \,E(\k)\, f(\k) \,,
\eeq
where $g_0 = 2(N_c^2-1)$ is the degeneracy factor for gluons with $N_c$ colors ($g_0 = 16$ for SU(3)). Generalizing the result of \cite{Zwanziger:2004np}, the energy density of the fluid is given in the covariant form by the expression~\cite{Florkowski:2015rua}
\beq
\label{eq:EpsCov}
\varepsilon = \int \DK\,E(k \cdot u)\, f(x,k) \,,
\eeq
where $f(x,k)$ is the phase space distribution function and the integration measure $\DK$ is defined as
\beq
\int\DK \left( \ldots \right) \equiv g_0 \int \frac{\dd^3k}{(2\pi)^3 k^0} \, \ku \left(\ldots \right)\,.
\label{DK}
\eeq
Here, the dispersion relation $E(k \cdot u)$ is given by Eq.~(\ref{eq:GZdispersionCov}). In the case of local thermal equilibrium, the gluon distribution has the Bose-Einstein form \cite{Zwanziger:2004np}
\beq
f_\GZ = \frac{1}{\exp\big[E(k\cdot u)/T(x) \big] - 1} \,,
\label{GZeqf}
\eeq
where the  temperature $T$ can depend on space and time. 

We proceed in a similiar fashion with the pressure, which in LRF is given by \cite{Zwanziger:2004np}
\begin{align}
P &= \frac{g_0}{3} \int \frac{\dd^3k}{(2\pi)^3} \,|\k| \frac{\partial E(\k)}{\partial |\k|} \, f_{\rm }(\k) \\
\label{eq:PressureNonCov}
&=  \frac{g_0}{3} \int \frac{\dd^3k}{(2\pi)^3} \frac{\k^2}{E(\k)} \left(1 - \frac{\gamma_\smallG^4}{\k^4} \right) \, f (\k) \,.
\end{align}
In the covariant version, the pressure reads~\cite{Florkowski:2015rua}
\beq
\label{eq:PressureTotCov}
P =  \frac{1}{3} \int \DK \,  \frac{(k\cdot u)^2}{E(k\cdot u )} 
\left[1 - \frac{\gamma_\smallG^4}{(k\cdot u)^4} \right]\, f (x,k)\,.
\eeq
The expressions for the energy density and pressure can be used to find the interaction measure (sometimes referred to as the trace anomaly)
\beq
\label{eq:InteractionMeasure}
I \equiv \varepsilon - 3 P = 2 \int \DK \frac{\gamma_\smallG^4}{(k\cdot u)^2 \, E(k\cdot u)} \, f(x,k) \,.
\eeq
This quantity is closely related to the non-conformality of the system under consideration. We note that $I$ vanishes identically for $\gamma_\smallG \to 0$ (or, more generally, in the conformal limit). 

Whenever the system is in local thermal equilibrium, the temperature dependence may be eliminated to construct the equation of state (EOS) of the GZ plasma
\beq
\label{eq:EOS}
\varepsilon_\GZ = \varepsilon_\GZ\left(P_\GZ \right) \,.
\eeq
This relation is further discussed in Sec.~\ref{sec:BjorkenHydro}.
In this case, the entropy density can be also calculated from the thermodynamic identity
\begin{eqnarray}
\label{eq:EntropyEq}
s_\GZ = \frac{\varepsilon_\GZ + P_\GZ}{T_\GZ} \,,
\end{eqnarray}
which is consistent with the conventional thermodynamic identities
\beq
\label{eq:ThermoId}
\dd\varepsilon_\GZ = T_\GZ \dd s_\GZ \,,\hspace{1em} \dd P_\GZ = s_\GZ \dd T_\GZ \,.
\eeq
In the opposite case, when the system is not in local equilibrium, one introduces an effective temperature of the system by the Landau matching condition
\begin{eqnarray}
\label{eq:LandauMatching0}
\varepsilon_\GZ(T) =\varepsilon \,.
\end{eqnarray}
In this case, the quantity $T$  is nothing else but a measure of the local, non-equilibrium energy density. 

\begin{figure}[t]
\centering
\includegraphics[width=0.95\columnwidth,clip=true,trim= 1mm 5mm 0mm 0mm]{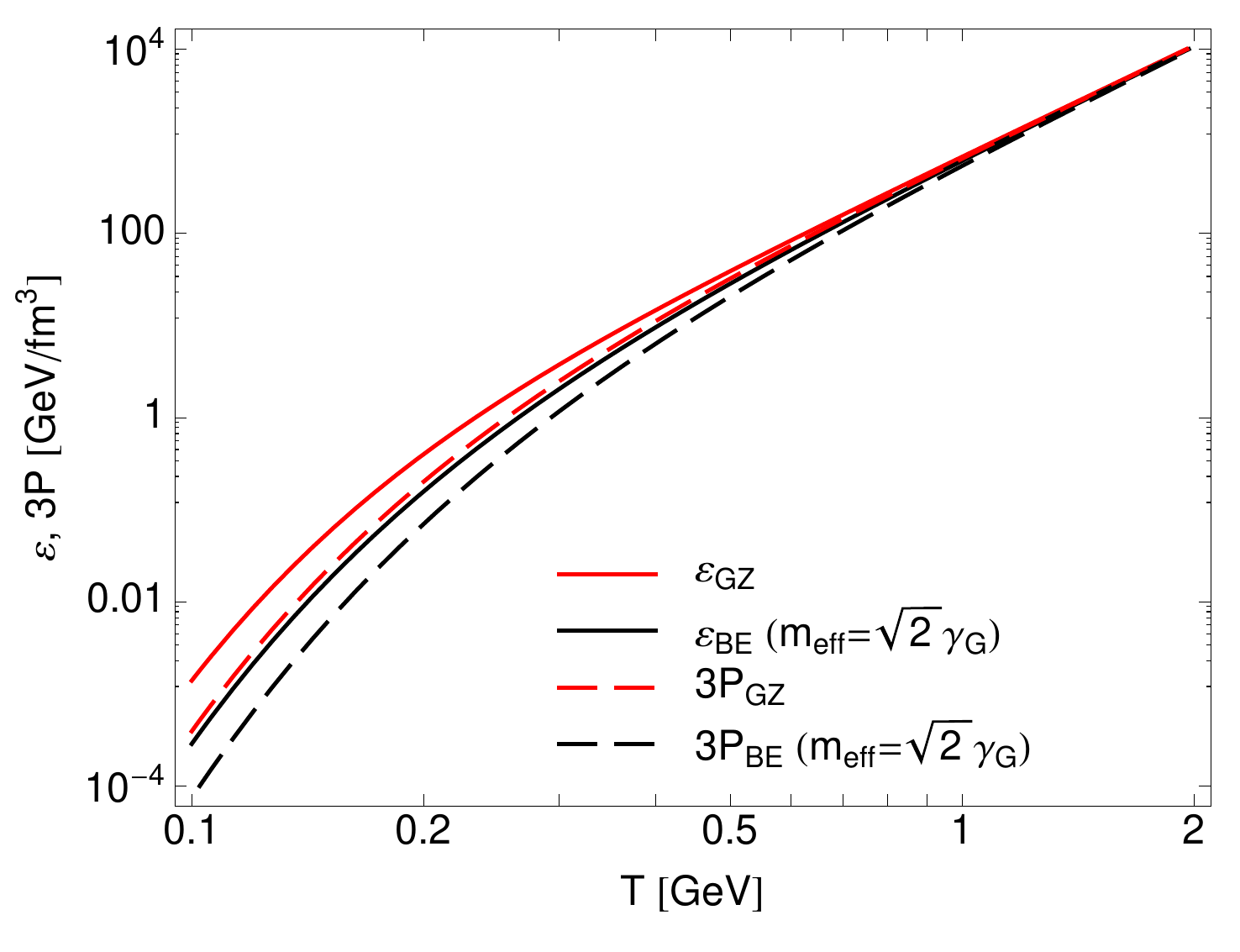}
\caption{\label{fig:EP_T} (Color online) The temperature dependence of the equilibrium energy density $\varepsilon$ and pressure $P$ (multiplied by a factor of 3) for the GZ plasma (red lines) and an ideal plasma of massive particles obeying BE statistics, labeled ``BE'' (black lines). In the latter case, the particles have the mass $m_{\rm eff} = \sqrt{2} \gamma_\smallG$.}
\end{figure}

The thermodynamics of the system is governed by the IR scale $\gamma_\smallG$. In the temperature range studied at RHIC and the LHC a systematic solution of the gap equation suggests $\gamma_\smallG \simeq {\rm const}$ \cite{Zwanziger:2004np}, so that the energy density is not simply proportional to $T^4$ \cite{Chernodub:2007rn,Lichtenegger:2008mh}. This is crucial for reproducing a  non-trivial peaked behavior of the energy density in this region \cite{Zwanziger:2004np,Fukushima:2013xsa}. As a matter of fact, in~\cite{Florkowski:2015rua} we used $\gamma_\smallG = 0.7$~GeV to describe the lattice data \cite{Borsanyi:2012ve} in the vicinity of the phase transition, in agreement with \cite{Zwanziger:2004np}. We stress that the SU(3) Yang-Mills theory with $T_c = $~260~MeV can be described only qualitatively in the temperature range $T \sim (1-5) T_c$ \cite{Zwanziger:2004np}. This is, nevertheless, expected since our simple approach does not include any sophisticated resummation scheme in the calculation of the gluon properties. For a more thorough study of the transition regime, $T\sim T_c$, one should also keep in mind the alterations of the Yang-Mills vacuum structure \cite{Kharzeev:2015xsa,Canfora:2015yia}. Finally, for temperatures much higher than the ones indicated above, one should keep in mind the relation of the Gribov scale to the YM magnetic scale that arizes from the gap equation \cite{Zwanziger:2006sc,Fukushima:2013xsa}. We leave possible further improvements of our model for future investigations. We only note here that the precise value of $\gamma_\smallG$ is not important for our present discussion. After fixing $\gamma_\smallG = 0.7$~GeV,  the characteristic temperature for phase transition in our model is about~$100$~MeV.

In Fig.~\ref{fig:EP_T} we show the temperature dependence of the energy density $\varepsilon_\GZ$ (red solid line) and pressure $P_\GZ$ (red dashed line) of the GZ plasma. For the temperatures exceeding the value of $\gamma_\smallG$, one observes that the GZ plasma behaves similarly to the ideal gas as the energy density equals three times the pressure \cite{Zwanziger:2004np}. For comparison, in Fig.~\ref{fig:EP_T} we have also shown the results for the energy density (black solid line) and pressure (black dashed line) of the massive BE plasma. It is natural to use the effective mass
\begin{eqnarray}
\label{masseff}
m_{\rm eff}  = \sqrt{2} \, \gamma_\smallG \,,
\end{eqnarray}
in the comparative calculations involving the standard massive case because it corresponds to the minimum of the Gribov dispersion relation for $|\k| = \gamma_\smallG$, where $E = \sqrt{2} \,\gamma_\smallG$. By inspection of the results presented in Fig.~\ref{fig:EP_T} we conclude, however, that the use of the effective mass, given by Eq.~(\ref{masseff}), is not sufficient to reconcile the results obtained for the GZ plasma and the massive BE plasma for moderate and low temperatures. The agreement at high temperatures follows from the fact that all particles may be treated as massless in this case. This is the Stefan-Boltzmann limit where we have $\varepsilon = 3P = 3c_\SB T^4$ and the entropy is $s_\SB = 4 c_\SB T^3$, with the Stefan-Boltzmann constant $c_\SB = g_0 \pi^2/90$. Further differences are discussed in Secs.~\ref{sec:BjorkenHydro} and \ref{sec:StrongCoupling}.

\subsection{Implementation of boost-invariance}

Below we consider a (0+1)-dimensional [(0+1)D], boost-invariant and transversally homogeneous system corresponding to the Bjorken model~\cite{Bjorken:1982qr}. It is described by the flow vector $u^\mu  = (t/\tau, 0,0, z/\tau)$. In this case, all scalar functions of space and time depend only on the longitudinal proper time $\tau$, defined as $\tau = \sqrt{t^2-z^2}$. One may furthermore introduce the boost-invariant variables \cite{Bialas:1987en}
\begin{align}
v &=  k^0 t - k_\parallel z  \,,\\
w &= k_\parallel t -  k^0 z \,.
\end{align}
The variables $v$ and $w$ replace $k^0$ and $k_\parallel$  --- for a boost-invariant system one can consider all the quantities in the plane $z=0$ where $\tau=t$,  $v = k^0 t$, and $w = k_\parallel t$. We find that
\begin{eqnarray}
k \cdot u = \frac{v}{\tau} = \sqrt{\frac{w^2}{\tau^2} + k_\perp^2},
\end{eqnarray}
so that the particle energy becomes a function of the boost-invariant variables $\tau$, $w$, and $k_\perp$, 
\begin{eqnarray}
E(k \cdot u) = E(\tau,w,k_\perp) = \sqrt{\frac{w^2}{\tau^2} + k_\perp^2  + \frac{\gamma_\smallG^4}{\frac{w^2}{\tau^2} + k_\perp^2} } .
\end{eqnarray}
The phase space integration measure can be written as
\beq
\int \DK (\ldots) = \frac{g_0}{(2\pi)^3} \int_{-\infty}^\infty \frac{\dd w}{\tau} \,\int \dd^2k_{\perp} (\ldots) \,.
\eeq
We note that the phase space distribution function, which is a Lorentz scalar, may depend only on $\tau$, $w$ and $k_\perp$, namely $f=f(\tau,w,k_\perp)$~\cite{Bialas:1987en}. 

For completeness, we also write down the expressions for the longitudinal and transverse pressure components that will be used in the following~\footnote{The parallel pressure  acts in the direction of the beam axis. The transverse pressure acts in the transverse direction to the beam and is the same for all such directions. }
\begin{align}
\hspace{-0.2em}P_\parallel &= \int \DK \,\frac{w^2}{\tau^2 E(\tau,w,k_\perp)}\left[1-\frac{\gamma_\smallG^4}{(w^2/\tau^2 + k_\perp^2)^2} \right]\!f, \\
\hspace{-0.2em}P_\perp &= \int \DK \, \frac{k_\perp^2}{2\, E(\tau,w,k_\perp)}\left[1-\frac{\gamma_\smallG^4}{(w^2/\tau^2 + k_\perp^2)^2} \right]\!f ,
\end{align}
such that the total pressure, $ P = (P_\parallel + 2 P_\perp)/3$, corresponds to Eq.~(\ref{eq:PressureTotCov}).

\section{First and zeroth order hydrodynamics for  (0+1)D systems}
\label{sect:firstord}

\subsection{Energy-momentum conservation}

The main dynamic equation governing the GZ plasma follows from the energy and momentum conservation. 
Using the energy density, $\varepsilon$, that is always kept equal to its equilibrium value by force of the Landau matching condition in Eq.~(\ref{eq:LandauMatching0}), for (0+1)D systems we can write
\beq
\label{eq:0+1Hydro0}
\frac{\dd \varepsilon}{\dd \tau} + \frac{\varepsilon + P_\parallel}{\tau} = 0 \,,
\eeq
see e.g. \cite{Muronga:2003ta,Baier:2006um}. Equation~(\ref{eq:0+1Hydro0}) can be further rewritten as
\beq
\label{eq:0+1Hydro}
\frac{\dd \varepsilon}{\dd \tau} + \frac{\varepsilon + P_\GZ}{\tau} - 
\frac{\frac{4}{3}\eta_{\rm eff} + \zeta_{\rm eff}}{\tau^2} = 0.
\eeq
Here we have expressed the shear and bulk viscous pressures \footnote{Generally, the shear tensor has 5 independent components. They are reduced to one independent variable $\pi$ for (0+1)D systems.},
\begin{align}
\label{smallpi}
\pi &= \frac{2}{3} \left(P_\perp-P_\parallel\right) \,, \\
\label{largepi}
\Pi &= P - P_\GZ \,,
\end{align}
in terms of their respective shear and bulk viscosities using relations valid at first order,
\begin{align}
\label{eq:ShearHydro}
\pi &= \frac{4}{3} \frac{\eta_{\rm eff}}{\tau} \,, \\
\label{eq:BulkHydro}
\Pi &= - \frac{\zeta_{\rm eff}}{\tau} \,,
\end{align}
where $P= (P_\parallel + 2 P_\perp)/3 $. For conformal systems the shear tensor is traceless, $\Pi~=~0$, which immediately signals a vanishing bulk viscosity.

Equation~(\ref{eq:0+1Hydro0}) can be treated in two different ways. In the first case it is treated as a consequence of the relativistic Boltzmann kinetic equation which yields $\varepsilon$, $P_\parallel$, and $P_\perp$ as functions of the proper time. Once $\varepsilon$, $P_\parallel$, and $P_\perp$, are known one can calculate the equilibrium pressure $P_\GZ$ (if the EOS is known, which is usually assumed) and determine the effective coefficients $\eta_{\rm eff}$ and $\zeta_{\rm eff}$ as functions of time.

In the second case, Eq.~(\ref{eq:0+1Hydro0}) is considered as one of the hydrodynamic equations, given by Eqs.~(\ref{eq:0+1Hydro0})--(\ref{eq:BulkHydro}), which can be solved only if the kinetic coefficients $\eta$ and $\zeta$ are known. Such equations are called the first-order (Navier-Stokes) fluid dynamic equations. In general, they are known to suffer from several deficiencies, but they usually represent a good description of systems that are very close to equilibrium. For such systems one determines the kinetic coefficients $\eta$ and $\zeta$ as functions of $T$. We skip the labels ``effective'' in this second case to stress that a different method of determination of the kinetic coefficients is used in this situation. Namely, instead of solving the kinetic equation for genuine non-equilibrium systems, one typically considers small perturbations around the equilibrium state to determine the linear response of the system. This method will be used to determine $\eta$ and $\zeta$ for the GZ plasma in Sec.~\ref{sect:kinco}.

The key point of our approach is that the two ways of treating the shear and bulk viscosities explained above should become equivalent for systems approaching local equilibrium. In Section ~\ref{sec:KinEq} we construct the kinetic equation and find its solutions. This allows us to determine  $\eta_{\rm eff}$ and $\zeta_{\rm eff}$ as functions of time. On the other hand, analyzing small perturbations of the equilibrium distributions in Sec.~\ref{sect:kinco} we find  $\eta$ and~$\zeta$ as functions of temperature. The consistency check that $\eta_{\rm eff}(\tau) = \eta(T(\tau))$ and $\zeta_{\rm eff}(\tau) = \zeta(T(\tau))$ demonstrates an overall consistency of our framework and supports our derivation of the functional forms for $\zeta(T)$ and $\eta(T)$.

\subsection{Bjorken hydrodynamics}
\label{sec:BjorkenHydro}

For perfect fluid (zeroth-order) hydrodynamics the transport coefficients vanish, $\zeta = \eta = 0$, and the (0+1)D system is described by the well known Bjorken evolution equation~\cite{Bjorken:1982qr},
\beq
\label{eq:Bjorken1}
\frac{\dd \varepsilon_\GZ(T(\tau))}{\dd\tau} + \frac{\varepsilon_\GZ(T(\tau)) +P_\GZ(T(\tau))}{\tau} = 0 \,.
\eeq
The solution of this equation will be denoted later as $T_\GZ(\tau)$. Using the thermodynamic relations in Eqs.~(\ref{eq:EntropyEq}) and (\ref{eq:ThermoId}), we find from Eq.~(\ref{eq:Bjorken1}) that
\begin{eqnarray}
\label{eq:BjorkenEntropy}
\frac{\dd s_{\GZ}}{\dd \tau} + \frac{s_\GZ}{\tau} = 0,
\end{eqnarray}
which has the scaling solution
\beq
s_\GZ(\tau) = s_{\GZ}(\tau_0) \frac{\tau_0}{\tau}.
\label{scalesol}
\eeq
This describes the characteristic drop of entropy density which is independent of the EOS of the system. The time dependence of the temperature in the Bjorken model $T_{\GZ}(\tau)$ is found from the expression
\beq
\label{eq:TemperatureGZ}
\frac{\dd\ln T_{\GZ}(\tau)}{\dd\ln \tau} = - c_s^2 \,,
\eeq
where
\beq
c_s^2 = \frac{\partial P_\GZ}{\partial \varepsilon_\GZ} =  \frac{\partial P_\GZ/\partial T_\GZ}{\partial \varepsilon_\GZ/\partial T_\GZ} 
\label{cs2}
\eeq
is the speed of sound of the GZ plasma. For conformal systems, e.g., a relativistic, massless gas, where $\varepsilon=3P$, we reproduce the well-known scaling solution $T(\tau) = T_0 (\tau_0/\tau)^{c_s^2}$ with $c_s^2 = \onethird$. For more general systems, Eq.~(\ref{eq:TemperatureGZ}) demonstrates that the evolution in local thermodynamic equilibrium is determined completely by the form of $c_s$.

\begin{figure}[t!]
\centering
\includegraphics[width=0.95\columnwidth,clip=true,trim= 0mm 0mm 0mm 0mm]{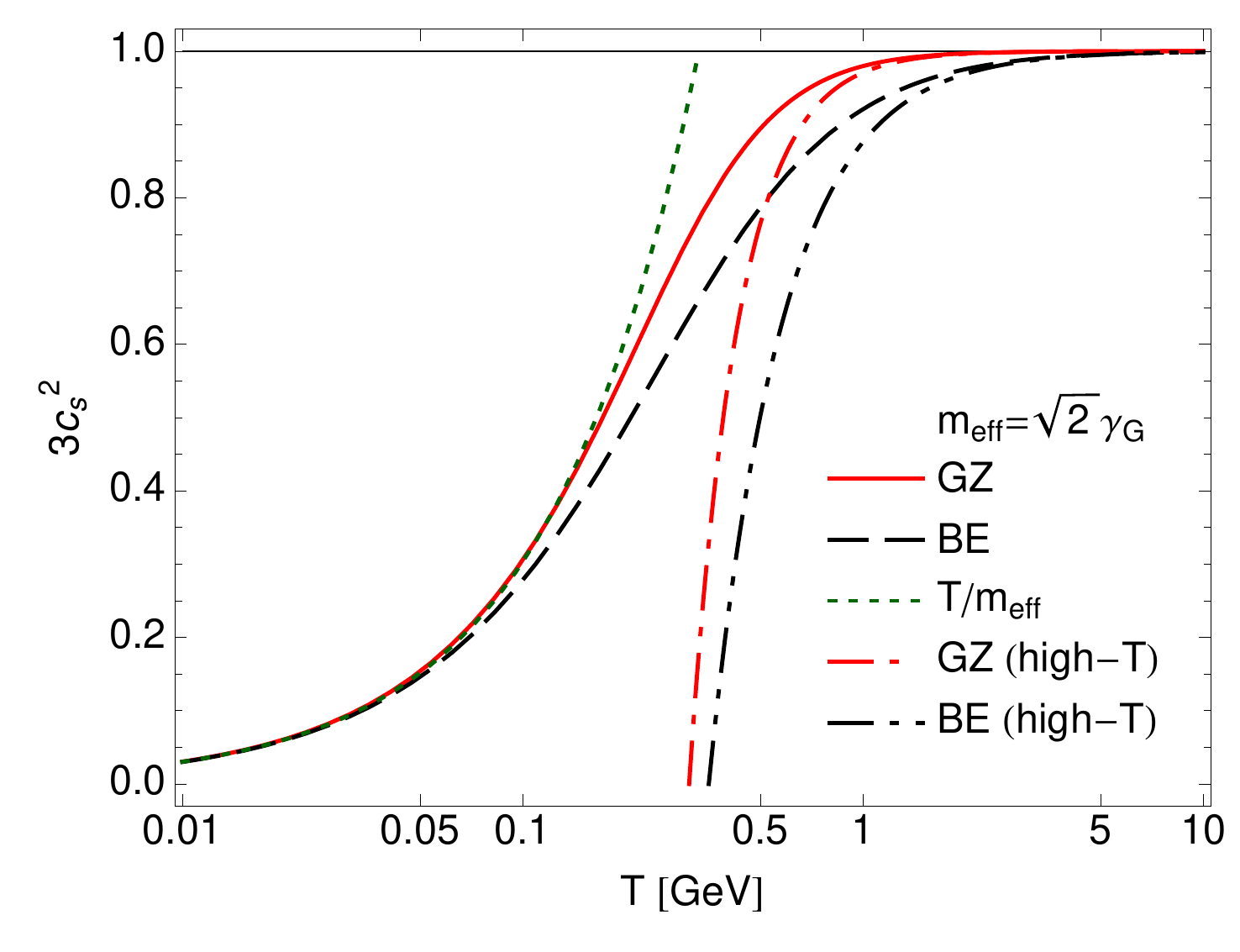}
\caption{\label{fig:cs2_T} (Color online) Temperature dependence of the sound velocity squared in the GZ plasma (red, solid line) and in the massive BE plasma (black, dashed line) with the particle mass given by Eq.~(\ref{masseff}). The green, dotted line represents the low-temperature limit \mbox{$c_s^2 = T/m_{\rm eff} = T/(\sqrt{2} \, \gamma_\smallG)$}, while the red, dotted line and the black, dashed-dotted line depict the first-order deviation at high-T for the two systems, respectively, see Eqs.~(\ref{eq:cs2highTGZ}) and (\ref{eq:cs2highTmassive}).}
\end{figure}

The temperature dependence of the sound velocity squared for the GZ plasma is shown in Fig.~\ref{fig:cs2_T} (solid red line). 
The result for the GZ plasma is compared with the result for the massive BE plasma (black dashed line). Interestingly, the two results differ for moderate and high values of $T$. At low temperatures the common limit for the two systems is (see \cite{Chojnacki:2007jc} and Appendix~\ref{sec:epsv})
\beq
c_s^2(T) = \frac{T}{m_{\rm eff}} = \frac{T}{\sqrt{2} \gamma_\smallG} \quad (T \to 0) \,.
\label{cs2lowt}
\eeq
The behavior at high temperatures differs notably. For the GZ plasma at $T\gg \gamma_\smallG$ we have~\cite{Zwanziger:2004np}
\beq
\label{eq:cs2highTGZ}
c_s^2 = \frac{1}{3}\left[1 - \frac{2}{3\sqrt{2} \pi \, c_\SB} \left(\frac{\gamma_\smallG}{T}\right)^3\right] + \mathcal{O}\left(\frac{1}{T^4} \right) \,,
\eeq
while for the BE plasma we have
\beq
\label{eq:cs2highTmassive}
c_s^2 = \frac{1}{3}\left[1 - \frac{6}{27\, c_\SB} \left(\frac{m}{T}\right)^2 \right] + \mathcal{O}\left(\frac{1}{T^3} \right) \,,
\eeq
see Appendix~\ref{sec:MassiveGas}. The high-$T$ expansion breaks down for $T  \lesssim \gamma_\smallG$ or $T  \lesssim m_{\rm eff}$ in the two cases, as seen in Fig.~\ref{fig:cs2_T}. The GZ plasma approaches the Stefan-Boltzmann limit $c_s^2=\onethird$ much faster than the BE plasma, since the deviations go like the third rather than the second power of the inverse temperature. The difference between the two systems under consideration is however mostly pronounced in the intermediate temperature regime, specifically for $T\sim \gamma_\smallG$, where the GZ plasma has a more pronounced change of behavior. These features reflect a harder EOS of the latter compared to the BE plasma and result in a more rapid change of the energy density and pressure around the transition temperature, in line with expectations from the lattice \cite{Borsanyi:2012ve}. The full, numerical result for the speed of sound is however crucial in order to describe the temperature regime relevant for present-day colliders, where both the low- and high-$T$ expansions fail, see Fig.~\ref{fig:cs2_T}.

\section{Kinetic equation in the relaxation-time approximation}
\label{sec:KinEq}

\subsection{Relaxation time approximation}
\label{sec:RTA}

Our main starting point for kinetic considerations is the formula for the energy density (\ref{eq:EpsCov}), which in the (0+1)D case depends only on proper time, $\varepsilon(\tau) = \int \DK E(\tau,w,k_\perp) f(\tau,w,k_\perp)$.
In order to study the time evolution, we take the derivative of the energy density with respect to the proper time, which gives
\begin{align}
\label{eq:dEps2}
&\frac{\dd \varepsilon}{\dd \tau}  + \frac{\varepsilon(\tau) + P_\parallel(\tau)}{\tau} \nonumber\\
&\hspace{1.5em} = \int \DK \, E(\tau,w,k_\perp) \, \frac{\partial f(\tau,w,k_\perp)}{\partial \tau} \,.
\end{align}
Here we have identified the term proportional to the longitudinal pressure. We point out that additional terms would appear in Eq.~(\ref{eq:dEps2}) if the Gribov parameter $\gamma_\smallG$ is allowed to be medium dependent.

The terms on the left-hand side of Eq.~(\ref{eq:dEps2}) should vanish due to energy-momentum conservation (\ref{eq:0+1Hydro0}) in a (0+1)D system. Thus,  the term  on the right-hand side of Eq.~(\ref{eq:dEps2}) should vanish as well. This suggests that we can use the standard kinetic equation in the relaxation-time approximation (RTA) of the form~\cite{Bhatnagar:1954zz,Baym:1984np,Baym:1985tna}
\begin{eqnarray}
\label{eq:KinEq}
\frac{\partial f(\tau,w,k_\perp)}{\partial \tau} 
= \frac{f_{\GZ}(\tau,w,k_\perp) - f(\tau,w,k_\perp)}{\trel (\tau)},
\end{eqnarray}
where
\begin{align}
\label{eq:LandauMatching1}
&\int \DK \, E(\tau,w,k_\perp) \, f_\GZ(\tau,w,k_\perp) \nonumber\\
&\hspace{1.5em} = \int \DK \, E(\tau,w,k_\perp) \, f(\tau,w,k_\perp) \,. 
\end{align}
In Eq.~(\ref{eq:LandauMatching1}) we recognize the Landau matching condition for the energy density, see Eq.~(\ref{eq:LandauMatching0}). 

At this point it is  important to emphasize that although Eqs.~(\ref{eq:KinEq}) and (\ref{eq:LandauMatching1}) are sufficient to construct a consistent (albeit simple) kinetic theory scheme, they are limited to the case of a time-independent relaxation time. In particular,  Eq.~(\ref{eq:LandauMatching1}) does not hold for a momentum-dependent relaxation time. For that to be the case, the relaxation time should be included in the integrands in~Eq.~(\ref{eq:LandauMatching1}).  The incorporation of the momentum dependent relaxation time would make our results more realistic by implementing the fact that particles with larger (smaller) momenta interact more weakly (strongly).  However, such a generalization would make our considerations of exact solutions of the kinetic equation much more complicated (not to speak about the fact that at the moment very little is known about the scattering of Gribov gluons). Nevertheless, although the present approach is restricted mainly to the case of a momentum-independent relaxation time, it can be applied easily in the situations where the relaxation time depends on the effective temperature of the system following Refs.~\cite{Florkowski:2013lya,Florkowski:2013lza}. Possible effects of inclusion of the momentum dependent relaxation time  in our framework are discussed in Sec.~\ref{sec:momdep} and in  Appendices \ref{sec:MomTauRel} and \ref{BEmomdep}.
 
Another simplification done implicitly in Eqs.~(\ref{eq:KinEq}) and (\ref{eq:LandauMatching1}) is that the relaxation time accounts mainly for inelastic collisions. This follows from the assumed form of the equilibrium distribution function (\ref{GZeqf}), where the chemical potential has been set equal to zero. If the system was dominated by elastic collisions, the number of particles would be (approximately) conserved  and the background equilibrium distribution (\ref{GZeqf}) would include a non-vanishing chemical potential  determined from an additional Landau matching condition (similar to Eq.~(\ref{eq:LandauMatching1})  but expressing the equality of particle densities rather than particle energies).

The formal solution of Eq.~(\ref{eq:KinEq}) with a time dependent relaxation time is~\cite{Florkowski:2013lza,Florkowski:2013lya,Florkowski:2014sfa}
\begin{align}
\label{formsol}
f(\tau,w,k_\perp) &= f_0(w,k_\perp) D(\tau,\tau_0) \nonumber\\
& + \int_{\tau_0}^\tau \, 
\frac{\dd\tau^\prime}{\trel(\tau')} D(\tau,\tau^\prime)
f_\GZ(\tau^\prime,w,k_\perp) \,,
\end{align}
where the damping function $D(\tau_2,\tau_1) $ has the form 
\begin{eqnarray}
D(\tau_2,\tau_1)  = \exp\left[-\int_{\tau_1}^{\tau_2} \frac{\dd \tau}{\trel(\tau) } \right].
\end{eqnarray}
If the relaxation time is extremely short, the form of Eq.~(\ref{eq:KinEq}) guarantees that the actual distribution function is in practice always equal to the equilibrium one. 

In order to construct the solution of Eq.~(\ref{formsol}) we have to know the dependence of $T$ and $\trel$ on the proper time $\tau$. We also have to know the initial distribution $ f_0(w,k_\perp)$, which for this work will be chosen simply as the isotropic GZ equilibrium distribution as given in Eq.~(\ref{GZeqf}). The  proper-time dependence of temperature is found self-consistently from the Landau matching condition, see Eq.~(\ref{eq:LandauMatchinEq}).

Although the relaxation time $\trel$ in (\ref{formsol}) may depend on time, in this work we fix it to be a constant in order to single out basic features of the modified IR dynamics. In the following, we present results for a range of relaxation times, 0.5~fm/c~$<~\trel~<$~2~fm/c, which are in the ballpark of phenomenologically relevant values~\cite{Florkowski:2013lya}. In order to avoid being biased by this choice we have also calculated the ratio of the bulk to the shear viscosities, $\zeta/\eta$, which is independent of the relaxation time and contains information about the dynamics of the system that is independent of the RTA scheme, see Sec. \ref{sect:kinco}. One should note, however, that the RTA itself allows for such independence.

The non-equilibrium evolution equations for the energy density, the longitudinal, transverse and total pressures, $\mathcal{I} = \{\varepsilon,P_\parallel, P_\perp, P\}$, respectively, are given by the universal formula
\begin{align}
\label{eq:EvolutionEq}
\mathcal{I}(\tau) &= D(\tau,\tau_0) H_\mathcal{I}\left(\frac{\gamma_\smallG}{T(\tau_0)},\frac{\tau_0}{\tau} \right) \nonumber\\
&+ \int_{\tau_0}^\tau \frac{\dd \tau'}{\tau_{\rm rel}(\tau')}D(\tau,\tau') H_\mathcal{I}\left(\frac{\gamma_\smallG}{T(\tau')},\frac{\tau'}{\tau} \right) \,,
\end{align}
where the auxiliary functions $H_{\mathcal{I}}$ are listed in Appendix~\ref{sect:integrals}. We choose the initial condition to be $T_0\equiv T(\tau_0) = 0.6$ GeV and $\tau_0 =$ 0.5 fm/c, as in \cite{Florkowski:2015rua}.

As stated above, the temperature of the system is fixed by the Landau matching condition which makes sure that the actual distribution function $f(\tau,w,k_\perp)$ yields the same energy density as the GZ equilibrium function $f_\GZ(\tau,w,k_\perp)$. This gives the implicit equation~\cite{Florkowski:2013lza,Florkowski:2013lya,Florkowski:2014sfa}
\begin{align}
\label{eq:LandauMatchinEq}
&H_\Eps\left(\frac{\gamma_\smallG}{T(\tau)},1 \right) = D(\tau,\tau_0) H_\Eps\left(\frac{\gamma_\smallG}{T(\tau_0)},\frac{\tau_0}{\tau} \right) \nonumber\\
&\hspace{1.5em} + \int_{\tau_0}^\tau \frac{\dd \tau'}{\tau_{\rm rel}(\tau')}D(\tau,\tau') H_\Eps \left(\frac{\gamma_\smallG}{T(\tau')},\frac{\tau'}{\tau} \right) \,,
\end{align}
which has to be solved for the temperature $T$ at the proper time~$\tau$, $T(\tau)$. Alternatively, in equilibrium, the proper-time dependence of the temperature can be found directly from Eq.~(\ref{eq:TemperatureGZ}), which corresponds to taking $\trel \to 0$ in Eq.~(\ref{eq:LandauMatchinEq}).

\subsection{Numerical results}
\label{sec:Results}

\begin{figure}[t]
\centering
\subfloat[]{\label{fig:TemperatureA}
\includegraphics[width=\linewidth,clip=true,trim= 0mm 0mm 0mm 2mm]{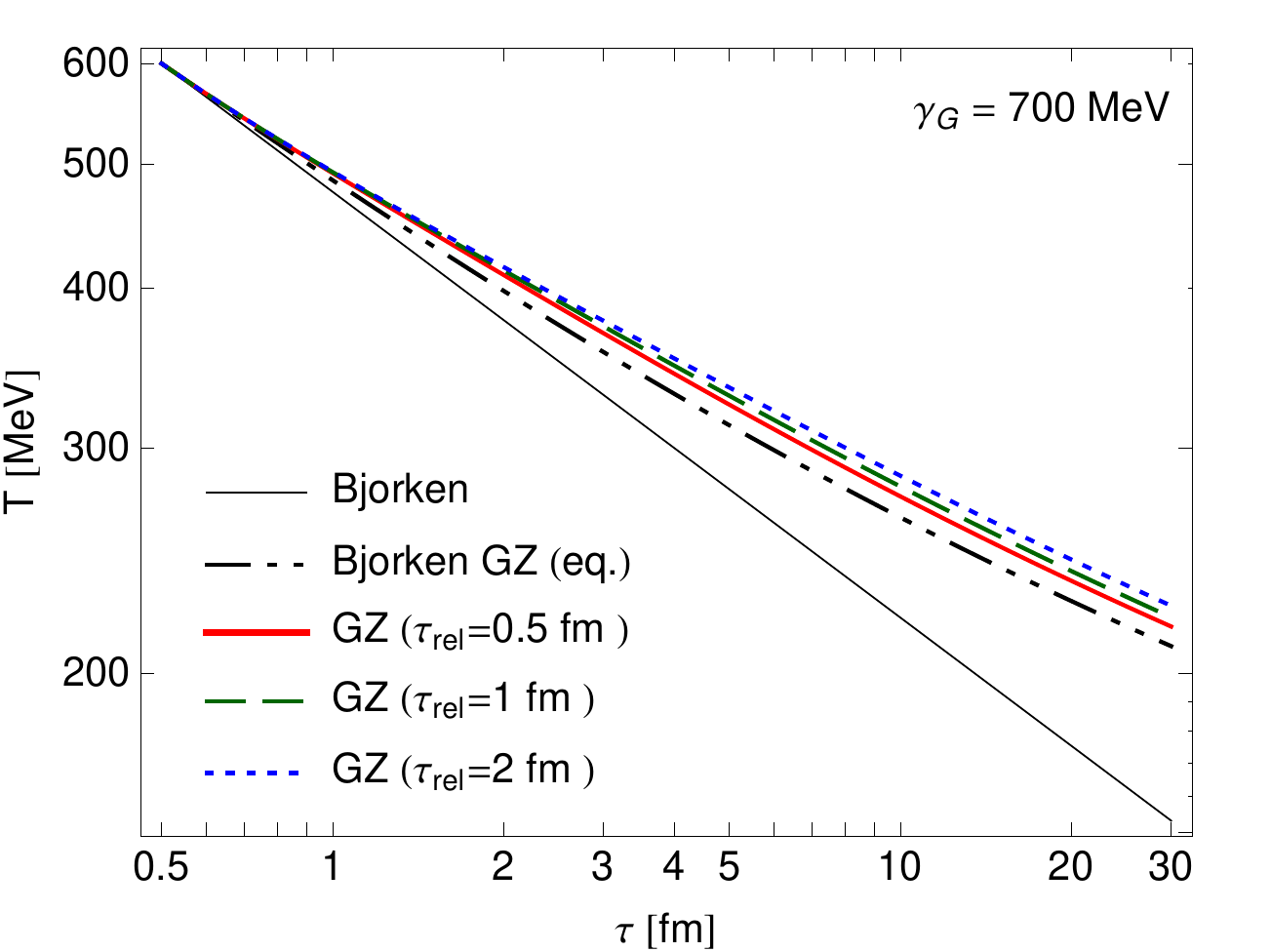}
}\\
\subfloat[]{\label{fig:TemperatureB}
\includegraphics[width=0.95\linewidth,clip=true,trim= 5mm 5mm 1mm 0mm]{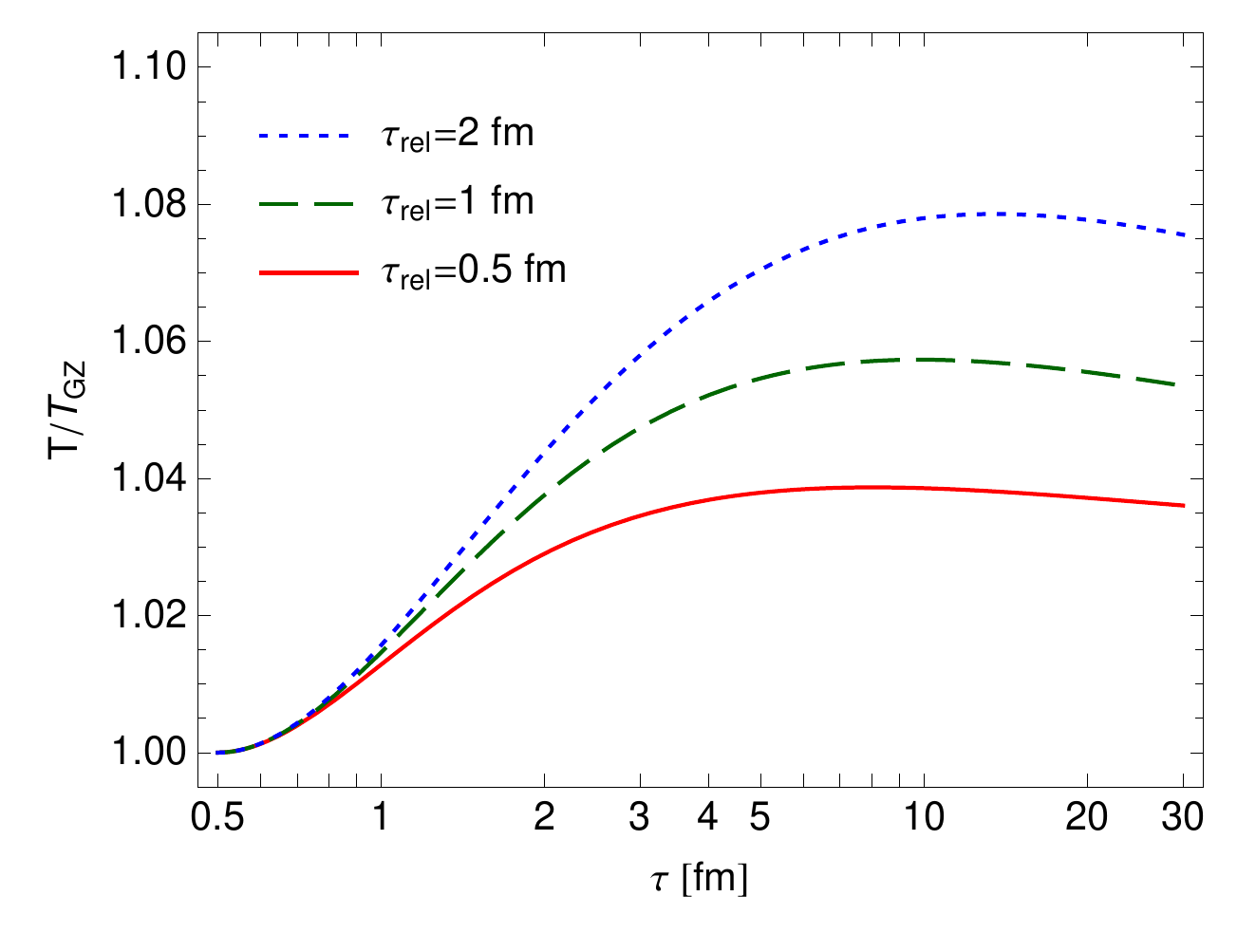}
}
\caption{\label{fig:Temperature} (Color online) The proper-time dependence of the effective temperature $T$ (left panel) and the same normalized by the equilibrium temperature $T_\GZ$ (right panel) for three values of the relaxation time, $\trel =$ 0.5~fm/c~(red, solid), 1~fm/c~(green, dashed) and 2~fm/c~(blue, dotted), respectively. The equilibrium temperature in the GZ plasma is plotted as the dashed-double-dotted line in the left plot, where we also have added the temperature dependence for an relativistic, massless gas (black, thin, solid line). We will keep these plotting conventions for all subsequent figures.}
\end{figure}

In Fig.~\ref{fig:TemperatureA} we plot the proper time dependence of the temperature for a massless, relativistic gas (black, solid, thin line), for the GZ plasma in equilibrium (black, dashed-double-dotted line) and for the GZ plasma with three chosen values of the relaxation time, $\trel = $ 0.5, 1 and 2 fm/c. In the lower panel, Fig.~\ref{fig:TemperatureB}, we plot the ratio of the effective temperature to the equilibrium temperature, $T/T_\GZ$, for the same choice of the relaxation times. With increasing relaxation time we find slower temperature evolutions and larger $T/T_\GZ$ ratios. This effect is due to increased entropy production within the system caused by dissipation.

\begin{figure}[t]
\centering
\includegraphics[width=0.95\columnwidth,clip=true,trim= 5mm 5mm 1mm 0mm]{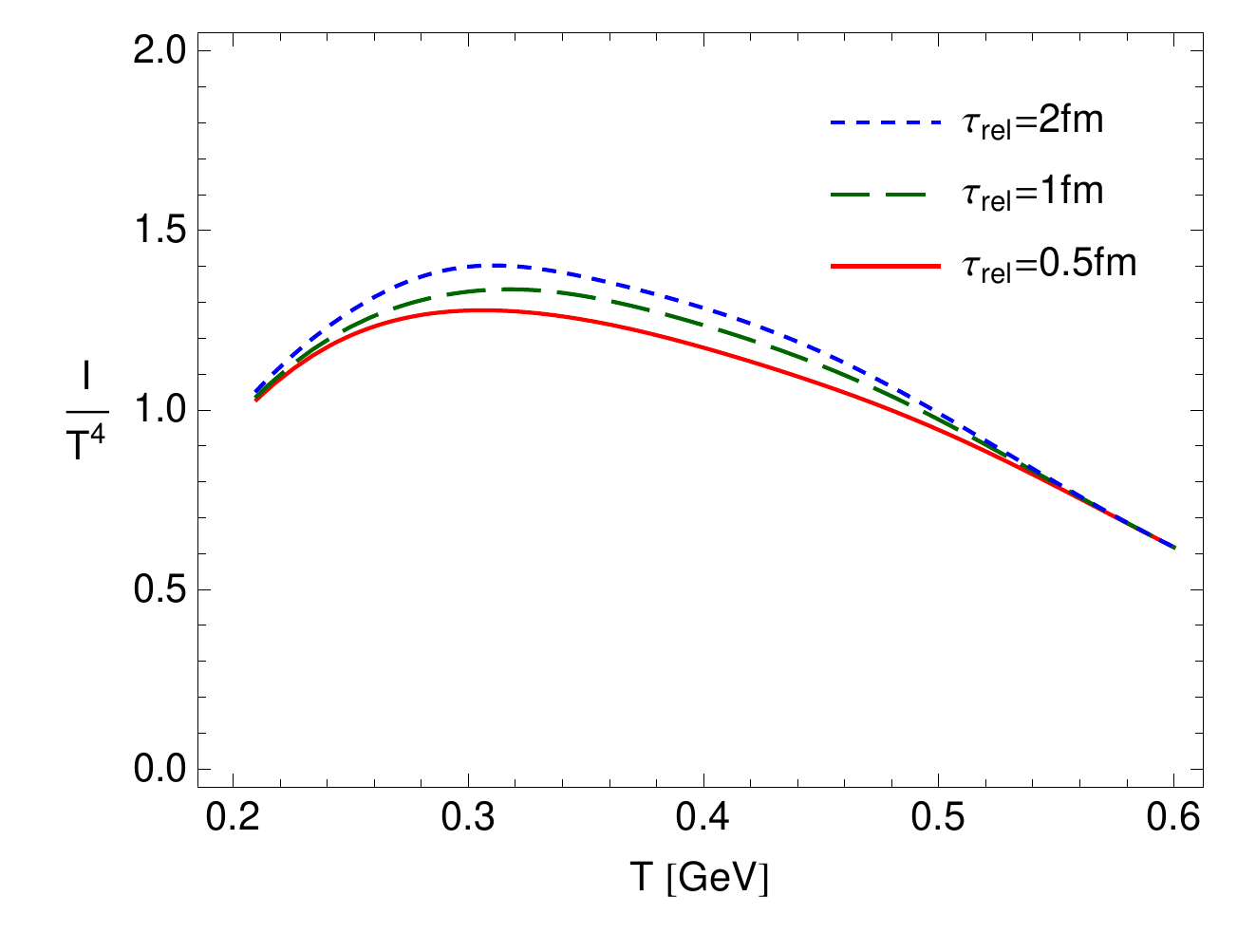}
\caption{\label{fig:InteractionMeas} (Color online) The interaction measure, Eq.~(\ref{eq:InteractionMeasure}), scaled by the fourth power of temperature, $I/T^4$, as a function of the effective temperature. The conventions for the labelling of the curves adopted from Fig.~\ref{fig:Temperature}.}%
\end{figure}

In Fig.~\ref{fig:InteractionMeas} the interaction measure~(\ref{eq:InteractionMeasure}) scaled by the fourth power of temperature, $I/T^4$, is plotted versus $T$. The three curves corresponding to the three different relaxation times agree at high and low temperatures. The agreement at high temperature follows from using the same initial condition in the kinetic equation. The agreement at low temperatures is a consequence of the fact that the analyzed systems are all evolving towards local equilibrium for sufficiently large proper time. Indeed, to cool the system from the initial temperature $T_0=600$ MeV down to $T=200$ MeV takes more than 20 fm/c, which is much larger than the largest relaxation time used in the calculations. Interestingly, the non-equilibrium evolution makes the interaction measure always larger than the equilibrium case \cite{Florkowski:2015rua}.

\begin{figure}[t]
\centering
\includegraphics[width=0.95\columnwidth,clip=true,trim= 1mm 5mm 1mm 0mm]{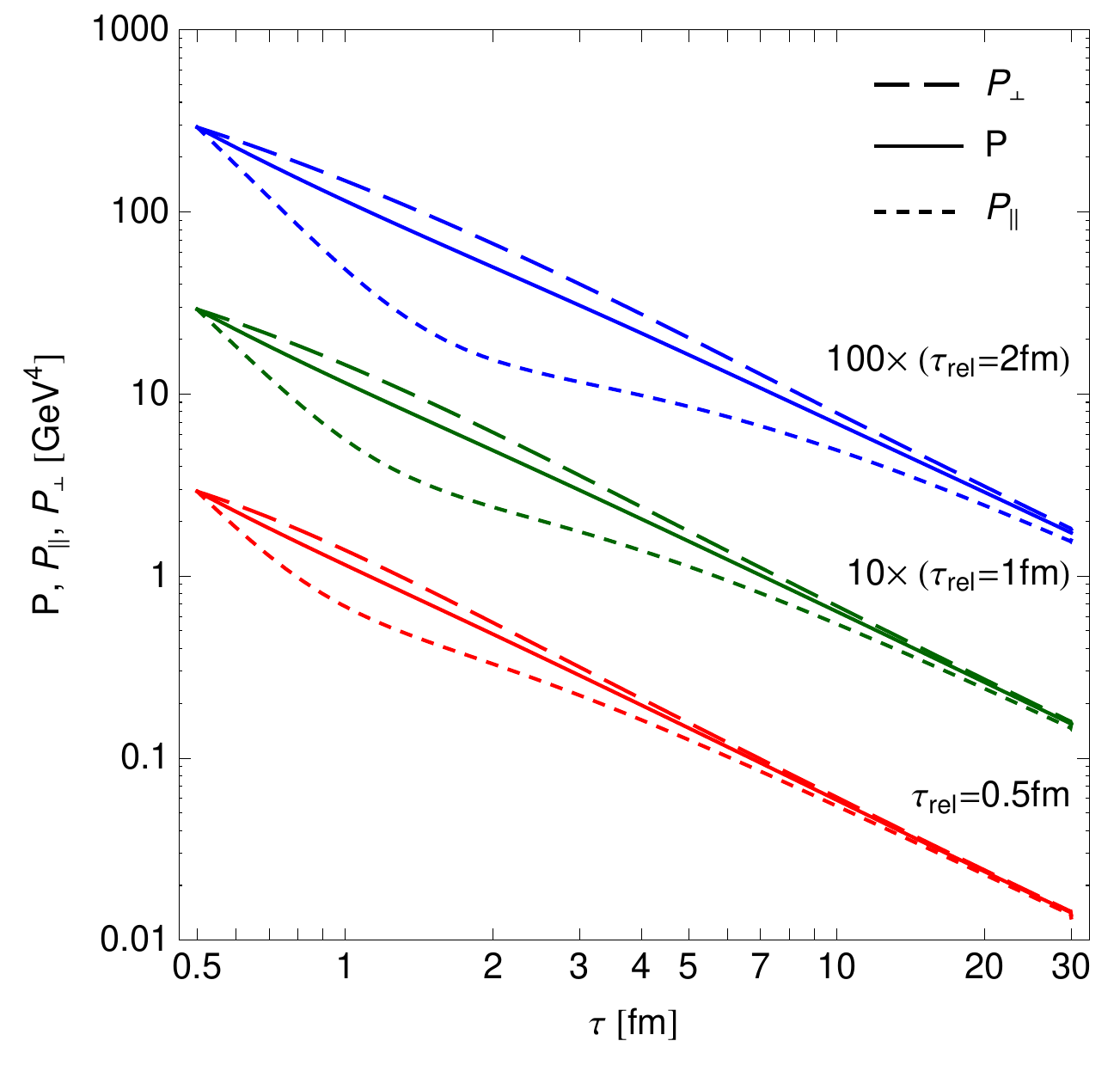}
\caption{\label{fig:PressuresTau} (Color online) Evolution of the various pressure components in time: the total (solid lines), longitudinal (dotted lines) and transverse pressure (dashed lines), respectively. The results are presented for three values of the relaxation time. The color coding is the same as in Fig.~\ref{fig:Temperature}.}%
\end{figure}

Dynamic properties of the GZ plasma are further described in Fig.~\ref{fig:PressuresTau} where we plot the proper time evolution of the total, longitudinal and transverse pressures. The results are presented again for three values of the relaxation time. In all the cases we observe an initial decrease of the longitudinal pressure and increase of the transverse pressure. Usually, this effect is explained by the presence of the shear viscosity that, due to strong longitudinal expansion, decreases the longitudinal pressure and increases the transverse pressure~\cite{Muronga:2003ta}. The differences in pressure become more pronounced with increasing viscosity, hence with the relaxation time. This is also seen in our calculations. We note, however, that the results in Fig.~\ref{fig:PressuresTau} represent exact solutions, cf. Eq.~(\ref{eq:EvolutionEq}), of the kinetic equation and do not necessarily rely on the interpretation related to the use of first-order dissipative fluid dynamics. We also note that the pressure components become approximately equal at a relatively late stage of the evolution of the system. In contrast, the agreement between the full kinetic theory and first-order dissipative fluid dynamics sets in relatively fast, approximately at $\tau \sim (3-5)\, \trel$, see Sec.~\ref{sect:kinco}.

\section{Transport coefficients}
\label{sect:kinco}

The basic non-equilibrium properties of the GZ plasma at late times are characterized by transport coefficients. In the absence of any conserved charges, the most important ones are the bulk and shear viscosities, see Eq.~(\ref{eq:0+1Hydro}). In this Section we recall our results on the bulk viscosity $\zeta$, obtained in \cite{Florkowski:2015rua}, and present new results on the shear viscosity $\eta$ and the ratio $\zeta/\eta$. In all the cases we first introduce the effective coefficients using the solutions of the kinetic equation (\ref{eq:KinEq}). Then, small perturbations around the equilibrium are studied in order to derive formulas for the transport coefficients that define $\zeta$ and $\eta$ as functions of the temperature $T$ and the Gribov parameter $\gamma_\smallG$. Numerical results, based on Eq.~(\ref{eq:LandauMatchinEq}), demonstrate the consistency of the two approaches whenever the system is close to equilibrium.

\subsection{Bulk viscosity}

Although the energy density is always equal to the equilibrium energy density (due to the Landau matching condition), this does not hold for the pressure. The difference between the actual and the equilibrium pressures describes the bulk viscous pressure, see Eq.~(\ref{largepi}). Using the definition of the trace anomaly, Eq.~(\ref{eq:InteractionMeasure}), and the Landau matching condition, Eq.~(\ref{eq:LandauMatching0}), one finds that
\begin{eqnarray}
\label{eq:BulkViscEff1}
\Pi = -\frac{2}{3} \int\DK \frac{\gamma_\smallG^4}{ (k\cdot u)^2 \,E(k\cdot u)} 
\left( f - f_{\GZ}\right) \,,
\end{eqnarray}
which simply corresponds to $(I^\GZ-I)/3$.
Knowing $\Pi$ from the kinetic theory, we may extract the effective bulk viscosity using Eq.~(\ref{eq:BulkHydro}), which gives the effective bulk viscosity coefficient, 
\beq
\zeta_\eff(\tau)  = - \tau \,\Pi(\tau).
\label{zetaeff}
\eeq
Alternatively, we may seek the solution of the kinetic equation (\ref{eq:KinEq}) in the form
\beq
\label{eq:LinExp}
f \approx f_\GZ + \delta f + \cdots \,,
\eeq
where $ \delta f = -\trel \dd f_\GZ/\dd \tau$ and the ellipsis represents higher order terms. We truncate the series at the linearized level in order to study the late-time behavior of the system. In this case,
\begin{align}
\label{eq:Deltaf}
\delta f &= -\frac{E \,\trel}{T \tau}\Bigg\{ \frac{w^2}{E^2 \tau^2} \left[1 - \frac{\gamma_\smallG^4}{\left(\frac{w^2}{\tau^2}+k_\perp^2 \right)^2} \right] \nonumber\\
& + \frac{\dd \ln T}{\dd \ln \tau} \Bigg\} f_{\GZ} \left(1+f_{\GZ}\right) \,.
\end{align}
The second term in the curly brackets can be replaced by the speed of sound, given in Eq.~(\ref{eq:TemperatureGZ}). Substituting $f - f_{\GZ} = \delta f$ in Eq.~(\ref{eq:BulkViscEff1}) and extracting the bulk viscosity from Eq.~(\ref{zetaeff}), yields a closed expression,
\beq
\label{eq:ZetaSimplified}
\hspace{-0.7cm} \zeta = \frac{g_0\gamma_\smallG^5 }{3\pi^2} \frac{\tau_{\rm rel}}{T}  \int_0^\infty \dd y \,\left[ \vsound^2 - \frac{1}{3}\frac{y^4 - 1}{y^4+ 1} \right] f_\GZ(1+f_\GZ),
\eeq
where $f_\GZ = \{\exp[\gamma_\smallG\sqrt{y^2+y^{-2}}/T]-1 \}^{-1}$ and $y~=~(k~\cdot~u)/\gamma_\smallG$. We have omitted the subscript ``eff'' since Eq.~(\ref{eq:ZetaSimplified}) results from the standard definition of the bulk viscosity that applies to first-order hydrodynamics.

\begin{figure}[t]
\centering
\includegraphics[width=0.95\columnwidth,clip=true,trim= 5mm 5mm 1mm 0mm]{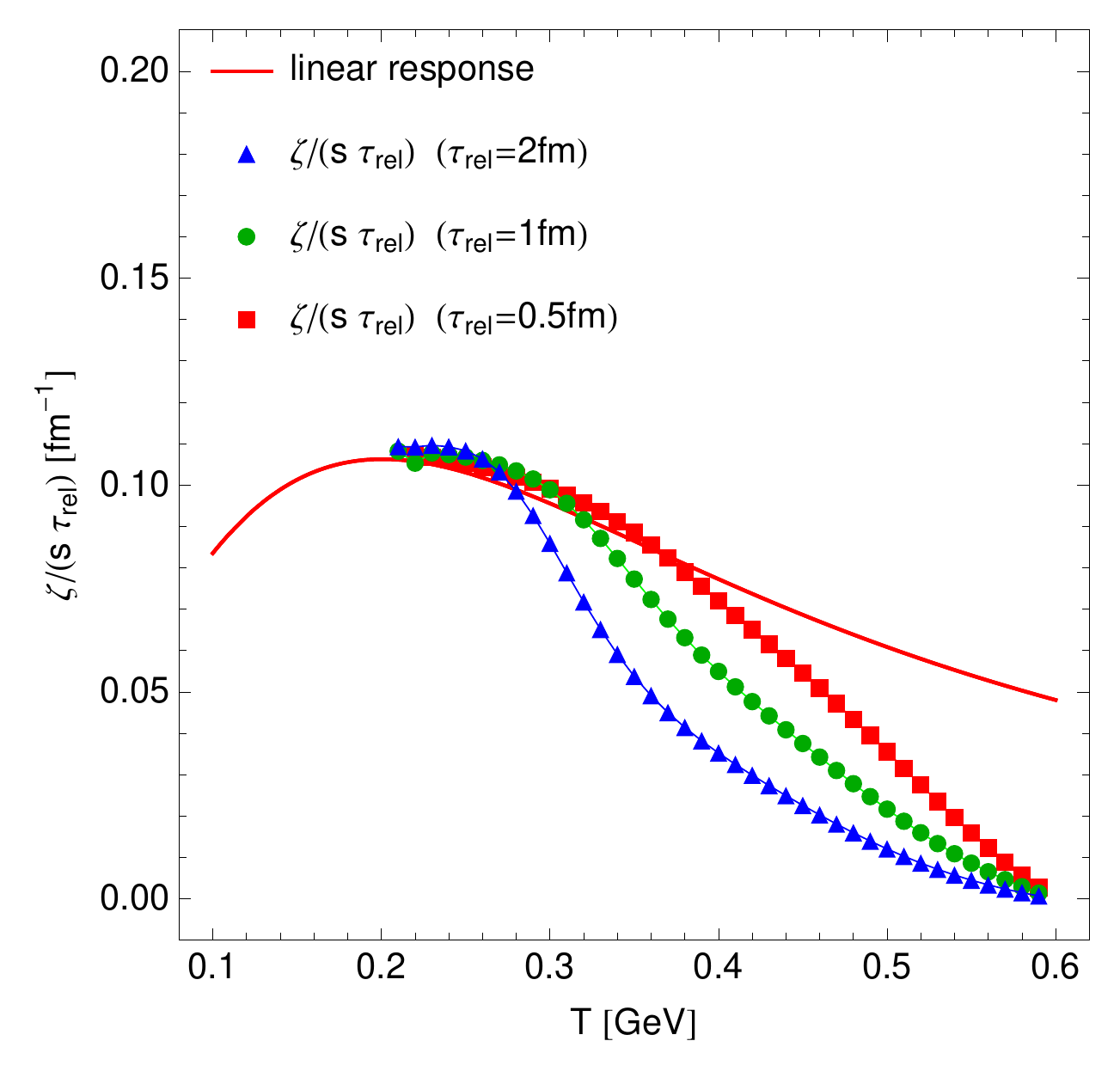}
\caption{\label{fig:BulkVisc} (Color online) Effective bulk viscosity scaled by entropy density and the relaxation time as a function of the temperature for three relaxation times: $\trel =$ 0.5 fm/c (red squares), $\trel =$ 1 fm/c (green circles) and $\trel = $2 fm/c (blue triangles). Analytic results in the linearized approximations (red solid line) are calculated using Eq. (\ref{eq:ZetaSimplified}). }%
\end{figure}

In Fig.~\ref{fig:BulkVisc} we plot the bulk viscosity obtained numerically from Eq.~(\ref{zetaeff}) and scaled by the entropy density and the relaxation time $\trel$, $\zeta/(s\trel)$, versus $T$ (filled symbols). Since we are interested in the properties of the system close to equilibrium, we simply approximate the a priori non-equilibrium entropy $s$ by the relation $(\varepsilon + P)/T$ (which is strictly only valid in local, thermal equilibrium). The results are compared with the linear-response result (red solid line), from Eq.~(\ref{eq:ZetaSimplified}). We observe a good agreement between the two calculations at low temperature, that correspond to large evolution times (much larger than the relaxation time). By comparing the results obtained with three different relaxation times, see Fig.~\ref{fig:TemperatureA}, we estimate the time when the system complies with a description in terms of linearized transport coefficients as $\sim5\,\trel$. For all three relaxation times, the ratio $\zeta/(s \trel)$ is non-monotonic and exhibits a maximum at a similar temperature as the maximum of the interaction measure, cf. Fig~\ref{fig:InteractionMeas}.

It is also instructive to determine the high-$T$ dependence of the ratio $\zeta/s$. In order to find the leading behavior it is simply sufficient to substitute the entropy density and the speed of sound by their respective values in the Stefan-Boltzmann limit, i.e. $s \approx s_\SB$ and $c_s^2 \approx \onethird$. In this case, i.e., for $T \gg \gamma_\smallG$ (with constant $\gamma_\smallG$ and $\trel$), we obtain
\beq
\label{eq:zetashighT}
\frac{\zeta}{s} = \frac{5}{8\sqrt{2}\pi^3} \frac{\gamma_\smallG^3 \trel}{T^2} + \mathcal{O}\left(\frac{1}{T^4}\right) \,.
\eeq

\subsection{Shear viscosity}
\label{sec:ShearViscosity}

The shear viscous pressure arises in locally anisotropic systems. In (0+1)D systems, this is measured by $\pi$, defined by Eq.~(\ref{smallpi}). From Eq.~(\ref{eq:ShearHydro}), we find that
\beq
\label{eq:ShearViscEff}
\eta_\eff = -\frac{3 \, \tau\,\pi}{4} = - \frac{\tau}{2} \left(P_\parallel-P_\perp \right)  \,.
\eeq
Whenever the system is close to equilibrium, we may again analyze small perturbations around the equilibrium distribution, which are described by Eqs.~(\ref{eq:LinExp}) and (\ref{eq:Deltaf}). The leading-order term $f_\GZ$ is isotropic and does not contribute to the difference $P_\parallel-P_\perp$. Similarly, the term $\dd \ln T/\dd\ln\tau$ in $\delta f$ is also isotropic and can be neglected. The remaining integral defining $\eta$ can be written in the form
\beq
\label{eq:EtaSimplified}
\eta =\frac{g_0\gamma_\smallG^5}{30\pi^2} \frac{\trel}{T} \int_0^\infty \dd y \,\frac{\left( y^4-1 \right)^2}{y^4+1} f_\GZ(1+f_\GZ) \,,
\eeq
where the definitions for $f_\GZ$ and $y$ are given below Eq.~(\ref{eq:ZetaSimplified}). Equation~(\ref{eq:EtaSimplified}) is one of the main new results presented in this work.

\begin{figure}[t]
\centering
\includegraphics[width=0.95\columnwidth,clip=true,trim= 5mm 5mm 1mm 0mm]{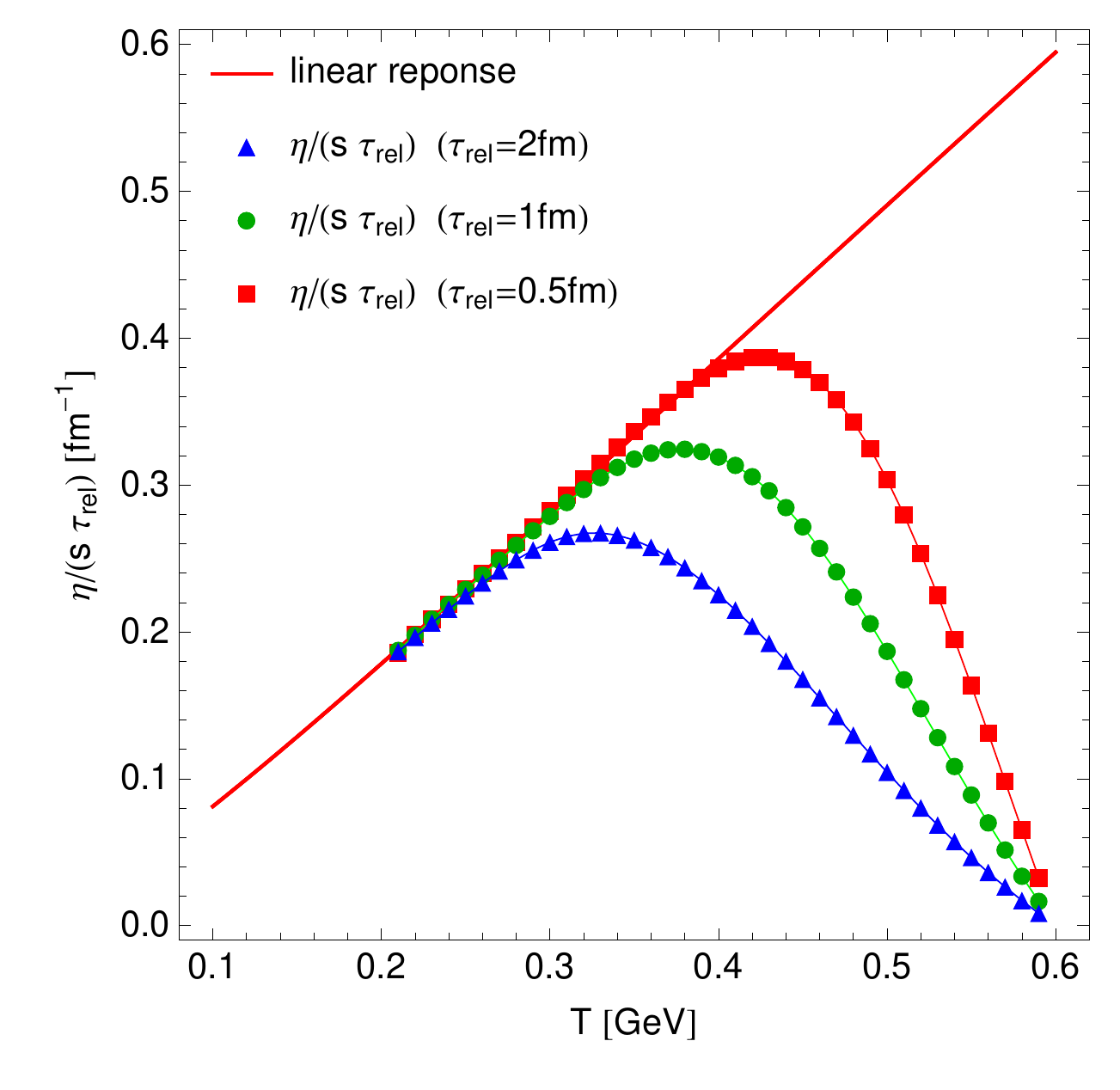}
\caption{\label{fig:SheaVisc} (Color online) The same as Fig.~\ref{fig:BulkVisc} but for the shear viscosity.}%
\end{figure}

In Fig.~\ref{fig:SheaVisc} we plot the shear viscosity to entropy density ratio, where $\eta$ was obtained numerically from Eq.~(\ref{eq:ShearViscEff}) and compare it with the result obtained using Eq.~(\ref{eq:EtaSimplified}). We employ the same approximation for the entropy density as in the previous subsection. The presentation of the results adopts the labeling conventions of Fig.~\ref{fig:BulkVisc}. We observe a very good agreement between the two calculations at proper times much larger than the relaxation time. While the linearized regime sets in slightly faster for the shear viscosity coefficient than for the bulk viscosity one, roughly at $3\,\trel$, there is an overall similar systematics for the two quantities.

In order to find the high-$T$ limit for the shear viscosity, it is useful to rewrite Eq.~(\ref{eq:EtaSimplified}) as 
\beq
\label{eq:EtaSimplified2}
\eta = \frac{g_0 \, \trel T^4}{30 \pi^2} \mathcal{K}\left( \frac{\gamma_\smallG}{T} \right) \,,
\eeq
where
\beq
\mathcal{K}(a) = \int_0^\infty \dd v \, \frac{\left(v^4 - a^4 \right)^2}{v^4+a^4} \frac{e^A}{\left(e^A - 1 \right)^2} \,,
\label{calK}
\eeq
where $v = \gamma_\smallG y/T$ and $A = \sqrt{v^2 + a^4/v^2}$. For asymptotic temperatures we simply pick up $\lim_{T\to \infty}\eta = \frac{g_0 \, \trel T^4}{30 \pi^2} \mathcal{K}(0)$, where $\mathcal{K}(0) = 4\pi^4/15$. Thus, since in the same limit the entropy of the GZ plasma simply goes to the Stefan-Boltzmann limit, we get that
\beq
\label{eq:etashighT}
\frac{\eta}{s} = \frac{\trel\, T}{5} +\mathcal{O}\left(\frac{1}{T} \right)\,.
\eeq
For a massless, relativistic ideal gas, this relation is exact at any temperature in the RTA \cite{Florkowski:2013lya}. Naturally, this limit also coincides with the result for the massive BE plasma, see Sec.~\ref{sec:MassiveGas}.

\subsection{$\zeta/\eta$ scaling}
\label{sec:StrongCoupling}

\begin{figure}[t]
\centering
\includegraphics[width=0.95\columnwidth,clip=true,trim= 0mm 0mm 0mm 0mm]{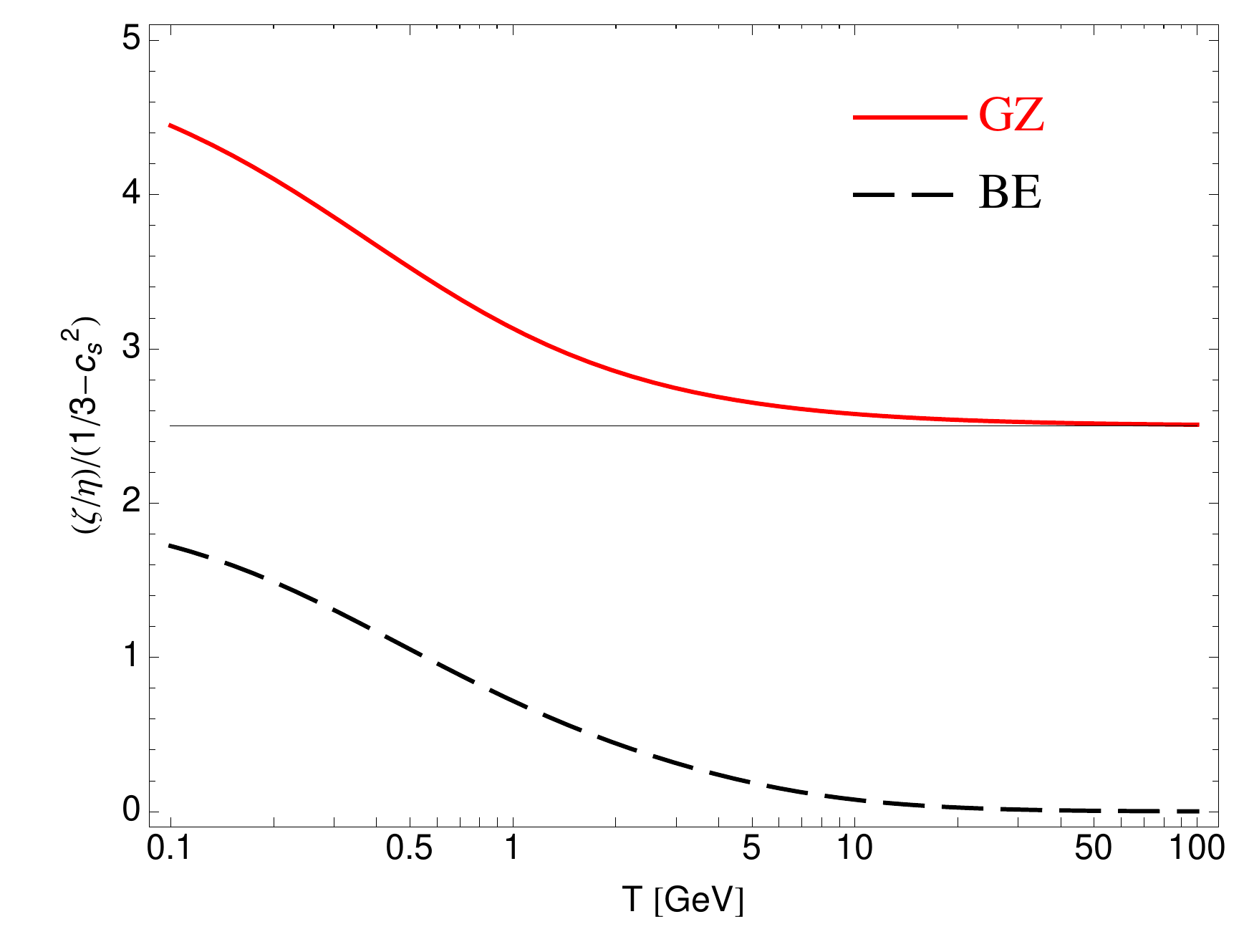}
\caption{\label{fig:ZetaEtaUniversality} (Color online) The ratio of the bulk and shear viscosities, $\zeta/\eta$, scaled by $\onethird - c_s^2$  for the GZ plasma (red, solid line) and the BE plasma (black, dashed line). The black, horizontal line indicates the value 5/2.}%
\end{figure}

From Eqs.~(\ref{eq:ZetaSimplified}) and (\ref{eq:EtaSimplified}), which were obtained assuming a constant Gribov parameter and relaxation time, we conclude that the ratio $\zeta/\eta$ is independent of the choice of the relaxation time. As a consequence, this ratio is a general prediction of our approach. Strikingly, while typical perturbative arguments predict \mbox{$\zeta/\eta \propto (c_s^2 -  \onethird)^2$} \cite{Weinberg:1971mx,Jeon:1995zm,Florkowski:2015lra}, we obtain a different form of the scaling. This is most easily demonstrated in the high temperature limit, $T \gg \gamma_\smallG$, where, combining Eqs.~(\ref{eq:zetashighT}) and (\ref{eq:etashighT}) which result in $\zeta/\eta \approx \frac{25}{8\sqrt{2}\pi^3} \left(\frac{\gamma_\smallG}{T} \right)^3$, with Eq.~(\ref{eq:cs2highTGZ}), gives
\beq
\label{eq:ZetaEtaRatio}
\frac{\zeta}{\eta} = \kappa_\GZ \left(\frac{1}{3} -c_s^2 \right) + \ldots \hspace{1em} (T\gg \gamma_\smallG)\,,
\eeq
with $\kappa_\GZ = 5/2$ and the ellipses stand for terms that are power-suppressed in the ratio $\gamma_\smallG/T$. In Fig.~\ref{fig:ZetaEtaUniversality} we plot $\zeta/\eta$ for the GZ plasma using Eqs.~(\ref{eq:ZetaSimplified})  and (\ref{eq:EtaSimplified}) (see the solid, red curve), where the limiting behavior is clearly seen at high temperature.  The linear scaling (\ref{eq:ZetaEtaRatio}) is characteristic for systems where the conformal symmetry is explicitly broken and has also been found in strongly coupled theories \cite{Benincasa:2005iv,Buchel:2005cv}. As an interesting observation, we note that the agreement between the Gribov-Zwanziger quantization at high temperature and strongly-coupled theories based on gauge-gravity duality is not a singular case, as the recently discovered massless mode of the QGP~\cite{Su:2014rma} is also closely in line with the holographic quasinormal mode~\cite{Kovtun:2005ev}.

It is instructive to contrast Eq.~(\ref{eq:ZetaEtaRatio}) with results for the BE plasma. The bulk viscosity of the BE plasma scales very similarly with $\frac{1}{3} -c_s^2$ as the result for the GZ plasma, see Eq.~(\ref{eq:ZetaMassive}) and the shear viscosity behaves as in the ideal gas at high temperature, see Eq.~(\ref{eq:zetashighT}). This is in agreement with the results found in \cite{Sasaki:2008fg}, which has also been derived within a quasiparticle model using RTA \cite{Huang:2010sa}. However, we find that the high-$T$ limit, $T \gg m_{\rm eff}$, results in $\frac{\zeta}{\eta} \approx  \frac{25}{16 \pi^3} \left(\frac{m_{\rm eff}}{T}\right)^3$, while the speed of sound deviates from its ideal value by a quadratic term, see Eq.~(\ref{eq:cs2highTmassive}). Therefore, we conclude that for the BE plasma at high temperatures, we find the relation
\beq
\frac{\zeta}{\eta} = \kappa_\BE \left(\frac{1}{3} - c_s^2 \right)^{3/2} + \ldots \hspace{1em} (T\gg m_{\rm eff}) \,,
\label{threehalvesscaling}
\eeq
with $\kappa_\BE = 3\sqrt{15}/2\approx 5.81$, which is qualitatively different from the linear relation found for the GZ plasma. The difference to the expected scaling with the second order \cite{Weinberg:1971mx,Jeon:1995zm,Florkowski:2015lra} can be traced to the lack of a momentum-dependent relaxation time as well as a temperature-dependent quasi-particle mass, as expected from perturbative arguments. We plot $\zeta/\eta$ for the BE plasma using Eqs.~(\ref{eq:ZetaMassive})  and (\ref{eq:EtaMassive}) (see the dashed, black curve) in Fig.~\ref{fig:ZetaEtaUniversality}. We note that the curve monotonically tends to zero at high temperature, confirming the absence of linear scaling.

\begin{figure}[t]
\centering
\includegraphics[width=0.95\columnwidth]{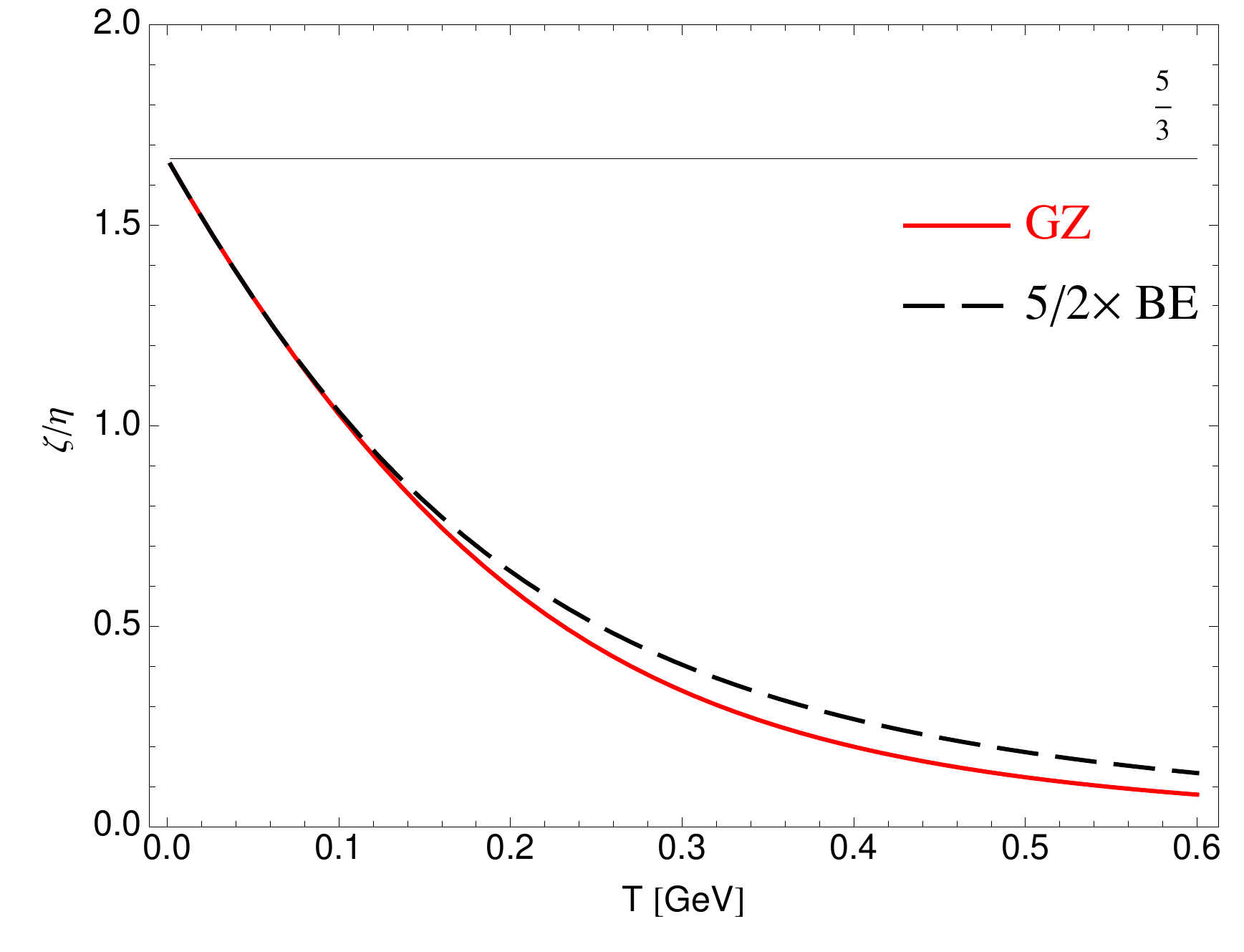}
\caption{\label{fig:zOe_T} (Color online) The temperature dependence of the bulk to shear viscosity ratio $n \times\zeta/\eta$ for the GZ plasma, see Eqs.~(\ref{eq:ZetaSimplified}) and (\ref{eq:EtaSimplified}) with $n = 1$ (red, solid line), and the massive BE plasma, see Eqs.~(\ref{eq:ZetaMassive}) and (\ref{eq:EtaMassive}) with $n = \frac{5}{2}$ (black, dashed line).}
\end{figure}

To further study the differences between the behavior of the GZ plasma and the massive BE plasma, in Fig.~\ref{fig:zOe_T} we plot the $\zeta/\eta$ ratio versus $T$ for these two systems. At low temperatures, $T \ll \gamma_\smallG$ or $T \ll m_{\rm}$, the $\zeta/\eta$ ratio tends to 5/3 for the GZ plasma and to 2/3 for the massive plasma, see Appendix~\ref{sec:LowTExpansionZetaEta} for the results of an analytic low-$T$ expansion. 
In terms of the constants in the respective scaling behaviors, this corresponds to $\kappa_\GZ = 5$ and $\kappa_\BE~=~2$, since $c_s^2 \to 0$ in both cases.
For a comparison of their respective low-$T$ behaviors, we have multiplied the BE result by 5/2 in Fig.~\ref{fig:zOe_T}. Surprisingly, the upper bound of $\zeta/\eta$ in the GZ plasma is in good agreement with the maximal value of this ratio from a recent gauge-gravity study \cite{Li:2014dsa}.

In order to access the phenomenologically relevant temperature regime we have to use the full expressions, Eqs.~(\ref{eq:ZetaSimplified}) and (\ref{eq:EtaSimplified}). In this regime, $T\sim(1-5)T_c$, the scaling relation in Eq.~$(\ref{eq:ZetaEtaRatio})$ is only approximate with a slowly varying $\kappa_\GZ$, see Fig.~\ref{fig:ZetaEtaUniversality}. We note, however, that the $\zeta/\eta$ ratio in the GZ plasma is always significantly larger than in the BE plasma. In particular, at small temperature $\zeta > \eta$ in the GZ plasma while this is never the case in the massive BE model.

\subsection{Comments on the momentum dependence of the relaxation time}
\label{sec:momdep}

Our main results have been obtained with two important assumptions, namely, that the Gribov parameter and the relaxation time are treated as constants. One may ask the question how these two assumptions affect our considerations. In order to analyze this issue, in Appendix~\ref{sec:MomTauRel} we relax the latter requirement presenting calculations done with a momentum-dependent relaxation time.  In this case, we assume a simple  power-law relation
\beq
\trel = \trel^0 (|\k|/\gamma_\smallG)^\alpha,
\label{eq:trelAlpha} 
\eeq
with $\alpha \geq 0$ being a parameter (for other discussions of the momentum dependent relaxation time see, for example, Refs.~\cite{Chakraborty:2010fr,Jaiswal:2016sfw}). 

The relaxation time (\ref{eq:trelAlpha}) should appear in the Landau matching condition (\ref{eq:LandauMatching1}) that had been used before to obtain Eqs.~(\ref{eq:BulkViscEff1}) and  (\ref{eq:ZetaSimplified}). With a more complex form of the relaxation time, we have to use a different strategy and employ (\ref{eq:Deltaf}) directly to calculate first the bulk viscous pressure and then to obtain the bulk viscosity. In this way,  instead of Eq.~(\ref{eq:ZetaSimplified}),  one finds the expression
\begin{align}
\label{eq:ZetaSimplifiedAlpha}
\hspace{-0.7cm} \zeta &= \frac{g_0\gamma_\smallG^5 }{6\pi^2} \frac{\trel^0}{T}  
\int_0^\infty  \dd y \,y^\alpha \,\, (y^4-1)
\nonumber \\
& \times \left[\frac{1}{3}\frac{y^4 - 1}{y^4+ 1} - \vsound^2  \right] f_\GZ(1+f_\GZ).
\end{align}
For the shear viscosity we can still use Eq.~(\ref{eq:EtaSimplified}), provided the relaxation time $\trel$ is moved into the integrand, which gives
\beq
\label{eq:EtaSimplifiedAlpha}
\eta =\frac{g_0\gamma_\smallG^5}{30\pi^2} \frac{\trel^0}{T} \int_0^\infty \dd y \,y^\alpha \frac{\left( y^4-1 \right)^2}{y^4+1} f_\GZ(1+f_\GZ) \,.
\eeq
We note that using the definition of the sound velocity one can check that in the case $\alpha=0$ Eq.~(\ref{eq:ZetaSimplifiedAlpha})  is equivalent to (\ref{eq:ZetaSimplified}) with $\trel = \trel^0$. 

\smallskip
Using  (\ref{eq:ZetaSimplifiedAlpha}) and (\ref{eq:EtaSimplifiedAlpha}) we construct the $\zeta/\eta$ ratio and find its asymptotic behaviour in the limit $\gamma_\smallG \ll T$. The analytic calculations presented in Appendix~\ref{sec:MomTauRel} show that for $\alpha \geq 1$ one gets
\beq
\frac{\zeta}{\eta} = 5 \left(\frac{1}{3}- c_s^2 \right).
\label{eq:scalingAlpha}
\eeq
For $0 < \alpha < 1$ one gets also a linear scaling with the factor 5 decreasing to 5/2 as $\alpha$ decreases from 1 to 0, which agrees with our previous result (\ref{eq:ZetaEtaRatio}). Thus we find that the scaling of the $\zeta/\eta$ ratio with the first power of the conformal measure turns out to be a universal feature of the GZ plasma.

In a recent paper, Jaiswal, Friman and Redlich showed that linear scaling holds also for a classical massive gas if the relaxation time depends linearly on the particle energy~\cite{Jaiswal:2016sfw}. In Appendix~\ref{sec:MassiveGas} we argue that our approach leads to identical results if applied to the same system. We find linear scaling also for the massive Bose-Einstein gas if the relaxation time depends on the momentum according to Eq.~(\ref{eq:trelBE}) with $\alpha > 0$. 
However, we note that these results have been obtained assuming temperature independent Gribov and mass parameters, in the case of the GZ and BE systems, respectively, which may affect the final form of the scaling.
We plan to analyze this issue in a separate work.

\section{Summary and conclusions}
\label{sec:Conclusions}

In this work we have continued our earlier investigations of the equilibrium and non-equilibrium properties of a plasma of gluons resulting from the Gribov-Zwanziger quantization (GZ plasma) in the relaxation time approximation. We have supplemented our earlier results for the bulk viscosity by the calculation of the shear viscosity and the $\zeta/\eta$ ratio as a function of temperature. We have demonstrated that the exact solutions of the kinetic equation support our expressions for the two viscosities. 

Several of the studied features suggest that at RHIC and LHC energies the GZ plasma is a strongly coupled system. Firstly, we observe a larger bulk viscosity, $\zeta/s$, compared to conventional quasiparticle approaches especially in the low temperature regime, where $\zeta > \eta$. This highlights its significant role for heavy-ion phenomenology. Secondly, for constant $\gamma_\smallG$ and $\trel$,  we find a linear scaling of the $\zeta/\eta$ ratio with $\onethird-c_s^2$, reflecting strong breaking of conformal symmetry of the system. At high temperature, this is at variance with the expected behavior of weakly-coupled plasmas \cite{Arnold:2006fz} and similar to expectations from a strongly-coupled system \cite{Benincasa:2005iv,Buchel:2005cv}.

Our results have been supplemented by the discussion of several aspects of the Gribov approach. In particular, we have analyzed the Lorentz covariance and found the low- and high-temperature expansions of thermodynamic functions and transport coefficients. We have also performed systematic comparisons between the qualitatively different properties of the Gribov-Zwanziger and the massive Bose-Einstein plasmas. In this exploratory study, our results have been obtained under the condition that the Gribov scale as well as the relaxation time are kept constant. Relaxing these assumptions allows to study more complex situations. 

Furthermore, we have checked the effects of a momentum-dependent relaxation time on our results and confirmed a linear scaling of the $\zeta/\eta$ ratio with the conformal measure also in this case. Similar scalings hold also for the massive Bose-Einstein and Boltzmann gas if the relaxation time is proportional to the momentum or energy of particles. This issue deserves further studies to clarify the interpretation of a linear scaling as a signal of strongly coupled systems.

In conclusion, we have demonstrated that the Gribov-Zwanziger quantization can be used to address in- and out-of-equilibrium physics in a unified way, which can be useful for future phenomenological applications to ultrarelativistic heavy-ion collisions. The latter may require the inclusion of a temperature-dependence of the Gribov scale. The first step in this direction has been done in Ref.~\cite{Begun:2016lgx}.

\begin{acknowledgments}

We thank Jorge Casalderrey-Solana, Maxim Chernodub, Mei Huang, Jan Pawlowski, Michael Strickland, Carsten Greiner and Zhe Xu for inspiring discussions. We also thank Amaresh Jaiswal for extended and clarifying discussions about the use of the momentum-dependent relaxation time. Research supported in part by Polish National Science Center grants No. DEC-2012/05/B/ST2/02528, No. DEC-2012/06/A/ST2/00390 (W.F.) and No. DEC-2012/07/D/ST2/02125 (R.R.). K.T. was supported by a Juan de la Cierva fellowship and by the Spanish MINECO under projects  FPA2013-46570 and 2014SGR104, partially by MDM-2014-0369 of ICCUB (Unidad de Excelencia 'Mar\'ia de Maeztu'), by the Consolider CPAN project and by FEDER. This research project has been supported by a Marie Sklodowska-Curie Individual Fellowship of the European Commission's Horizon 2020 Programme under contract number 655279 ``ResolvedJetsHIC''. N.S. was supported by the Helmholtz International Center for FAIR within the framework of the LOEWE program launched by the State of Hesse.

\end{acknowledgments}

\medskip
\appendix

\section{Implementation of Lorentz covariance}
\label{sect:cov} 

Gribov's dispersion relation, Eq.~(\ref{eq:GZdispersionNonCov}), relating the particle three-momentum $\k$ and energy $E$ explicitly breaks Lorentz invariance. The reason for this breaking is the use of the Coulomb gauge $\nabla \cdot {\bf A} = 0$ in the derivation of Eq.~(\ref{eq:GZdispersionNonCov}).  Without the lack of generality, we may assume that the Coulomb gauge is imposed in the inertial reference frame ${\cal S}$. The four-velocity of the whole system in  ${\cal S}$ is $u^{\mu} = (1,0,0,0)$, hence,  we rewrite  Eq.~(\ref{eq:GZdispersionNonCov}) as two combined equations
\beq
\label{eq:GZdispersionNonCov1}
E_k^2 - \k^2  =  \frac{\gamma_\smallG^4}{\k^2}, \quad u^{\mu}=(1,0,0,0).
\eeq
We may try to rewrite Eq.~(\ref{eq:GZdispersionNonCov1}) as a single covariant formula. The first natural choice seems to be the introduction of the four-vector ${\tilde k} = (E_k, \k)$ with the Minkowski square ${\tilde k}^2 = E_k^2 - \k^2$, then Eq.~(\ref{eq:GZdispersionNonCov1}) can be written covariantly as
\beq
\label{eq:GZdispersionNonCov2}
{\tilde k}^2  = \frac{\gamma_\smallG^4}{({\tilde k} \cdot u)^2 - {\tilde k}^2}.
\eeq
In the frame where $u^{\mu}=(1,0,0,0)$, Eq.~(\ref{eq:GZdispersionNonCov2}) is reduced naturally to (\ref{eq:GZdispersionNonCov1}). However, when (\ref{eq:GZdispersionNonCov2}) is considered in other frames, it typically yields four different solutions (some of them complex) for the energy $E_k$ determined at given values of $\k$ and ${\bf v}$. This leads to a paradox, since in  ${\cal S}$ all energy states are real and equal. 

Clearly, one should restrict oneself to the energies defined in the local rest frame of the fluid element. A natural way to achieve this and implement the covariant description is to consider $\k$ as the spatial part of the four vector $k = (k_0 = |\k|, \k)$ and to treat the energy $E$ defined in (\ref{eq:GZdispersionNonCov}) as a scalar
\beq
\label{eq:GZdispersionNonCov3}
E = \sqrt{ (k \cdot u)^2 + \frac{\gamma_\smallG^4}{(k \cdot u)^2}}.
\eeq
In the frame ${\cal S}$, where $u^{\mu} = (1,0,0,0)$,  Eq.~(\ref{eq:GZdispersionNonCov3}) is reduced directly to (\ref{eq:GZdispersionNonCov1}). In the frame  ${\cal S}^\prime$ we have
\beq
\label{eq:GZdispersionNonCov4}
E = \sqrt{ (k^\prime \cdot u^\prime)^2 + \frac{\gamma_\smallG^4}{(k^\prime \cdot u^\prime)^2}},
\eeq
where $u^\prime$ and $k^\prime$ are the four-velocity of the frame ${\cal S}$ and the four-momentum $k$ seen in the frame ${\cal S}^\prime$. For one dimensional systems considered in this work, the components of $u^\prime$ may be treated as the parameters of the Lorentz transformation $k^\prime \cdot u^\prime = k_0^\prime u_0^\prime - k_z^\prime u_z^\prime = k_0 = |\k|$. Hence, Eq.~(\ref{eq:GZdispersionNonCov3}) in frames other than ${\cal S}$ gives real and non-degenarate values of the energy, in agreement with the original definition of the energy done in the frame ${\cal S}$, where the Coulomb gauge was imposed. 

The prescription (\ref{eq:GZdispersionNonCov3}) has been adopted in this work. We note that our treatment of energy is similar to the treatment of temperature in the special theory of relativity. Temperature measures average kinetic energy and has been treated by many authors as the time component of the four-vector. However, since $T$ is measured in the LRF, its treatment as a scalar quantity seems to be more appropriate. The energies appearing in Gribov's formula have been obtained also in the special reference frame where the Coulomb gauge condition holds. This suggests that $E$  defined in such a way may be treated as a scalar quantity in the similar way as $T$.

\section{Auxiliary functions}
\label{sect:integrals} 

The energy density is calculated via the auxiliary function $H_\varepsilon(\gamma_\smallG,a, b)$, defined by
\beq
H_\varepsilon(a, b) = \frac{g_0 \gamma_\smallG^4}{2\pi^2} \int_0^\infty \dd y \frac{y \,h_\varepsilon(y,b)}{e^{a \sqrt{y^2 +  y^{-2}} } -1} ,
\label{Hene}
\eeq
with
\begin{eqnarray}
h_\varepsilon(y,b) = b \int_0^{\pi/2}\dd\varphi \sin\varphi \frac{\sqrt{y^4\beta^4 + 1}}{\beta} \,,
\end{eqnarray}
where $\beta^2 = b^2 \cos^2\varphi + \sin^2\varphi$, such that
\beq
h_{\varepsilon}(y,1) = \sqrt{ y^4 +1} \,.
\eeq
The energy density in equilibrium is given simply by
\beq
\Eps^\GZ(T) = H_\Eps(\gamma_\smallG/T,1).
\label{eqene}
\eeq

Next, we work out the auxiliary functions for the longitudinal and transverse pressures $\PL$ and $\PT$, respectively. They are from the expressions
\begin{align}
H_{P_\parallel}(a, b) &= \frac{g_0 \gamma_\smallG^4}{2\pi^2} \int_0^\infty \dd y \, \frac{y \,h_{\PL}(y,b)}{e^{a \sqrt{y^2 +  y^{-2}} } -1} \,,\\
H_{P_\perp}(a, b) &= \frac{1}{2}\frac{g_0 \gamma_\smallG^4}{2\pi^2} \int_0^\infty \dd y \, \frac{y \,h_{\PT}(y,b)}{e^{a \sqrt{y^2 +  y^{-2}} }-1} \,,
\end{align}
where
\begin{align}
h_{\PL}(y,b) &= b^3 \int_0^{\pi/2} \dd \varphi \frac{\sin\varphi \cos^2\varphi \left(y^4\beta^4 - 1 \right)}{\beta^3 \sqrt{y^4\beta^4 + 1}} \,,\\
h_{\PT}(y,b) &= b \int_0^{\pi/2} \dd \varphi \frac{\sin^3\varphi \left(y^4\beta^4 - 1 \right)}{\beta^3 \sqrt{y^4\beta^4 + 1}} \,,
\end{align}
such that
\begin{align}
h_{\PL}(y,1)& = \frac{1}{3}\frac{y^4 - 1}{\sqrt{y^4 + 1}} \,,\\
h_{\PT}(y,1) &= \frac{2}{3} \frac{y^4 - 1}{\sqrt{y^4 + 1}} \,.
\end{align}
The equilibrium pressure components are then $\PL^\GZ(T) = H_\PL(\gamma_\smallG/T,1)$ and $\PT^\GZ(T) = H_\PT(\gamma_\smallG/T,1)$, and it is straightforward to check that in this case $\PL^\GZ(T) =\PT^\GZ(T) $.

The total pressure is given by $\P = \frac{1}{3}(\PL + 2 \PT)$, where the auxiliary function responsible for its evolution is easily deduced to be
\beq
\label{Hpe}
H_{P}(a, b) = \frac{g_0 \gamma_\smallG^4}{6\pi^2} \int_0^\infty \dd y\frac{y \,h_{\P}(y,b)}{e^{a \sqrt{y^2 + y^{-2}} }-1} \,,
\eeq
with
\beq
h_{\P}(y,b) = b \int_0^{\pi/2} \dd \varphi \frac{\sin \varphi \left(y^4\beta^4 - 1 \right)}{\beta \sqrt{y^4\beta^4 + 1}} \,,
\eeq
such that
\beq
h_{\P}(y,1) = \frac{y^4 - 1}{\sqrt{y^4 + 1}} \,.
\eeq
The total pressure in equilibrium is then
\beq
\P^\GZ(T) = H_\P(\gamma_\smallG/T,1) \,.
\label{Pe}
\eeq

\section{Low temperature expansions}

In this Section we analyze different physical quantities, such as the equilibrium energy density, 
equilibrium pressure, the shear and bulk viscosity coefficients, and derive analytic expressions for 
all of them in a form of the series expansion with the leading terms corresponding to the 
low-temperature limit $a =\gamma_\smallG/T \gg 1$. By including in the systematic way
more terms in the expansions, we may obtain a successful description also in 
the region of moderate temperatures.

\subsection{Equilibrium energy density}
\label{sect:lowTene} 

We start with the equilibrium energy density. Combining our earlier results (\ref{Hene})--(\ref{eqene}) 
for the case $b=\beta=1$, we find
\beq
\Eps^\GZ = \frac{g_0 \gamma_\smallG^4}{2\pi^2} \int_0^\infty \dd y \, \frac{y \, 
\sqrt{ y^4 +1} }{e^{a \sqrt{y^2 + y^{-2}} }-1} \,.
\label{eqene0}
\eeq
By splitting this integral into two parts, the first from 0 to~1 and the second from 1 to $\infty$, and 
changing the integration variable $y$ to $1/y$ in the first part we get
\beq
\Eps^\GZ = \frac{g_0 \gamma_\smallG^4}{2\pi^2} \int_1^\infty
\frac{\dd y}{y}  \frac{(y^3 +y^{-3}) \, \sqrt{ y^2 +y^{-2}} }{e^{a \sqrt{y^2 + y^{-2}} }-1} \,.
\label{eqene1}
\eeq
In the next step we introduce the integration variable $z$ defined through the relations
\beq
y &=&  \sqrt{z^2 + \sqrt{z^4 - 1}}\,\,, \nonumber \\
\frac{\dd y}{y} &=& \frac{z~\dd z}{\sqrt{z^4-1}}.
\label{zvar}
\eeq
Then,  we can rewrite Eq.~(\ref{eqene1}) as
\beq
\Eps^\GZ = \frac{g_0 \gamma_\smallG^4}{2\pi^2} \int_1^\infty 
\dd z  \, \frac{ \fenez }{e^{ \s2 a z}-1} \,,
\label{eqenez}
\eeq
where the function $\fenez$ is defined through the formula
\beq
\fenez = \frac{\s2 z^2	}{\sqrt{z^4-1}} \left(y^3 + y^{-3} \right),
\label{fene}
\eeq
where $y$ is the function of $z$ defined in (\ref{zvar}). For $a \gg 1$, the main contribution 
to the integral (\ref{eqenez}) comes from the region $z \gtrsim 1$, where $\fenez$ has the expansion
\begin{align}
\fenez &= \frac{\s2}{\sqrt{z-1}} + \frac{23 \sqrt{z-1}}{2\s2 } + \frac{307 (z-1)^{3/2}}{16 \s2} \nonumber \\
&+ \frac{739 (z-1)^{5/2}}{64 \s2} + \frac{1667 (z-1)^{7/2}}{1024 \s2} + \cdots .
\label{feneexp1}
\end{align}
We represent this expansion in a more general form as
\beq
\fenez = \sum_{n=0}^\infty \aepsn (z-1)^{n-1/2},
\label{feneexp2}
\eeq
where the coefficients $\aepsn$ me be read off from (\ref{feneexp1}). In this way we find the expression 
for the equilibrium energy density as the following series
\beq
\Eps^\GZ = \frac{g_0 \gamma_\smallG^4}{2\pi^2} \sum_{n=0}^\infty \aepsn 
\int_1^\infty \dd z  \, \frac{ (z-1)^{n-1/2} }{e^{ \s2 a z}-1} \,.
\label{eqene2}
\eeq
In the next step it is convenient to change the integration variable $z$ to the variable $\alpha$ 
defined by equation $z = 1 + \alpha^2$. In this way we get
\beq
\Eps^\GZ = \frac{g_0 \gamma_\smallG^4}{\pi^2} \sum_{n=0}^\infty \aepsn 
\int_0^\infty  \dd \alpha\,  \frac{\alpha^{2n}}{e^{ \s2 a (1+\alpha^2) }-1} \,.
\label{eqene3}
\eeq
This integral over $\alpha$ can be done analytically and expressed by the gamma and polylogarithm functions, 
\begin{align}
\Eps^\GZ &= \frac{g_0 \gamma_\smallG^4}{2 \pi^2} \sum_{n=0}^\infty \aepsn 
\left(\s2 a\right)^{-(n+\frac{1}{2})} \nonumber\\
&\times \Gamma\left(n+\frac{1}{2}\right) \Lisn12 .
\label{EneGL}
\end{align}
At very low temperatures, $a = \gamma_\smallG/T  \gg 1$, the most important contribution is that with $n=0$. In this case we get
\beq
\Eps^\GZ \approx \frac{g_0 \gamma_\smallG^4  }{\s2 \pi^{3/2}} \frac{e^{-\s2 a}}{\left(\s2 a\right)^{1/2}}.
\label{EneGL0}
\eeq

\subsection{Equilibrium pressure and sound velocity}
\label{sec:epsv}

Starting from Eqs.~(\ref{Hpe})--(\ref{Pe}) and following the same steps as in the case of the equilibrium energy density one finds
\beq
\hspace{-2em} \P^\GZ = \frac{g_0 \gamma_\smallG^4}{6\pi^2} \int_0^\infty \dd y \frac{y^4 - 1}{\sqrt{y^2 + y^{-2}}}\frac{1}{e^{a \sqrt{y^2 + y^{-2}}}-1} ,
\label{pe0}
\eeq
and
\beq
\P^\GZ = \frac{g_0 \gamma_\smallG^4}{6\pi^2} \int_1^\infty 
\dd z \, \frac{\fpez }{e^{\s2 a z }-1} \,,
\label{pe1}
\eeq
where the function $\fpez$ is defined through the formula
\beq
\fpez = \frac{y^5 + y^{-5}-(y + y^{-1})}{\s2 \sqrt{z^4-1}} \,.
\label{fpe}
\eeq
In the region $z \gtrsim 1$,  $\fpez$ has the series expansion
\begin{align}
\fpez &=  6 \s2 \sqrt{z-1} + \frac{19 (z-1)^{3/2}}{\s2} \nonumber \\
& + \frac{93 (z-1)^{5/2}}{8 \s2} +  \frac{51 (z-1)^{7/2}}{32 \s2}+  \cdots\,,
\label{fpeexp1}
\end{align}
which defines the coefficients $\apen$ used in the formula
\beq
\P^\GZ = \frac{g_0 \gamma_\smallG^4}{6\pi^2} \sum_{n=1}^\infty \apen 
\int_1^\infty \dd z  \, \frac{ (z-1)^{n-1/2} }{e^{ \s2 a z } -1} \,.
\label{pe2}
\eeq
Note that $a_P^{(0)}=0$. The integration over $z$ can be done analytically in the same way 
as in the case of the energy density, and we find an analogous expression to Eq.~(\ref{EneGL}), namely
\begin{align}
\P^\GZ &= \frac{g_0 \gamma_\smallG^4}{6 \pi^2} \sum_{n=0}^\infty \apen 
 \left(\s2 a\right)^{-(n+\frac{1}{2})} \nonumber\\
 &\times \Gamma\left(n+\frac{1}{2}\right) \Lisn12 \, . 
\label{peGL}
\end{align}
At very low temperatures, $a = \gamma_\smallG/T  \gg 1$, the leading term in (\ref{peGL}) is that with $n=1$,
\beq
\P^\GZ \approx \frac{g_0 \gamma_\smallG^4  }{\s2 \pi^{3/2}} \frac{e^{-\s2 a}}{\left(\s2 a\right)^{3/2}}.
\label{peGL0}
\eeq

Equations~(\ref{EneGL}) and (\ref{peGL}) can be used to determine the sound velocity in the system. In the case of the temperature independent Gribov parameter $\gamma_\smallG$ we obtain
\beq
c_s^2 = \frac{\partial \P^\GZ}{\partial \Eps^\GZ} = \frac{
\partial \P^\GZ \big/ \partial a}{\partial \Eps^\GZ \big/ \partial a} \,.
\label{cs2ap1}
\eeq
At low temperatures we may use ~(\ref{EneGL0}) and (\ref{peGL0}) as the approximations for (\ref{EneGL}) and (\ref{peGL}). Keeping in mind that $a = \gamma_\smallG/T$, we find
\beq
c_s^2(T) =  \frac{1}{\s2 a} = \frac{1}{\s2} \frac{T}{\gamma_\smallG}.
\label{cs2ap2}
\eeq
We note that this result is similar to the result obtained for the massive plasma where we find $c_s^2 = T/m$ \cite{Chojnacki:2007jc}, where $m$ is the mass of particles in the plasma. 

\begin{figure}[t]
\centering
\includegraphics[width=0.95\columnwidth]{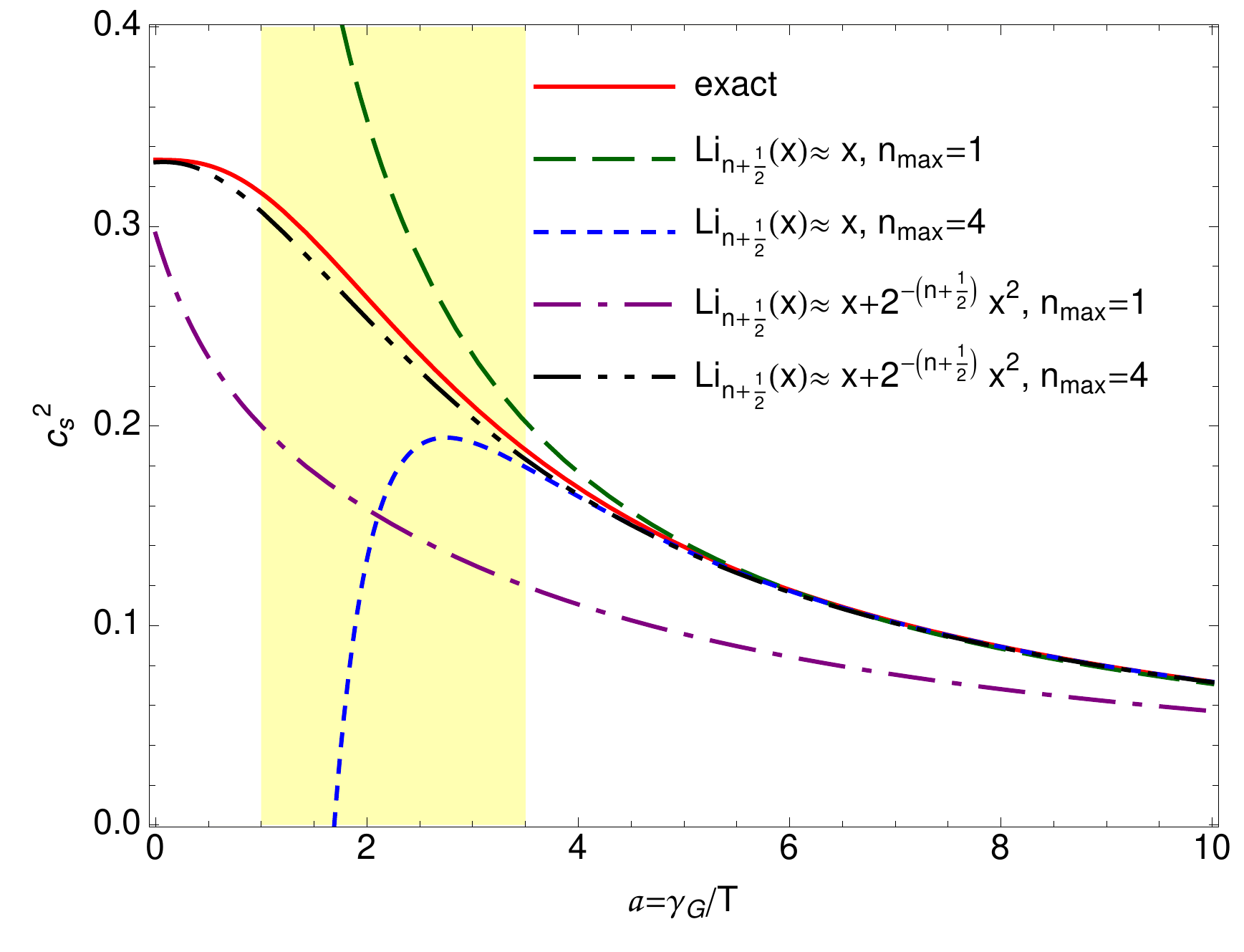}
\caption{\label{fig:cs2app} (Color online) The $a=\gamma_\smallG/T$ dependence of the exact speed of sound squared compared to approximated formula (\ref{cs24}) up to  $n=1$ (green dashed line)  and  $n=4$ (blue dotted line). For completeness we show corresponding results for $n=1$ (purple dashed-dotted line)  and  $n=4$ (black dashed-double-dotted line) obtained by including first NLO term in $\Lisn12$.}%
\end{figure}

One may go beyond the leading-order term in the expansions (\ref{EneGL}) and (\ref{peGL}) and include higher-order terms in $n$ in order to describe the regime close to the phase transition. Taking $\Lisn12 \approx e^{-\s2 a}$, up to $n = 4$, we get 
\begin{align}
\Eps^\GZ &\approx \frac{g_0 \gamma_\smallG^4  }{\s2 \pi^{3/2}} \frac{e^{-\s2 a}}{\left(\s2 a\right)^{1/2}} 
\left[
1
+\frac{23}{8}\left(\frac{1}{\s2 a}\right) 
\right.\nonumber\\
&+ \frac{921}{128}\left(\frac{1}{\s2 a}\right)^2 + \frac{11085}{1024}\left(\frac{1}{\s2 a}\right)^3  \nonumber\\
& +\left. \frac{175035}{32768}\left(\frac{1}{\s2 a}\right)^4 \right] \,,
\label{EneGL4}
\end{align}
and 
\begin{align}
\P^\GZ &\approx \frac{g_0 \gamma_\smallG^4  }{\s2 \pi^{3/2}} \frac{e^{-\s2 a}}{\left(\s2 a\right)^{3/2}} 
\left[1 + \frac{19}{8}\left(\frac{1}{\s2 a}\right) \right.\nonumber\\
& + \left. \frac{465}{128}\left(\frac{1}{\s2 a}\right)^2 
+ \frac{1785}{1024}\left(\frac{1}{\s2 a}\right)^3
\right] \,,
\label{peGL4}
\end{align}
and accordingly
\begin{align}
c_s^2(T) &=  \frac{1}{\s2 a} \left[1
+\frac{1}{2}\left(\frac{1}{\s2 a}\right) \right.\nonumber\\
&- \left.\frac{29}{8}\left(\frac{1}{\s2 a}\right)^2
-\frac{63}{8}\left(\frac{1}{\s2 a}\right)^3
 \right] \,.
\label{cs24}
\end{align}
In Fig.~\ref{fig:cs2app} we present the comparison of the approximated formula (\ref{cs24}) with the exact one. We observe that, as expected, the expansion (\ref{cs24}) works very well already at leading order in $n$ (dashed green line) if the temperatures are very small. However, at the same time, we see that the expansion completely fails at temperatures close to and above the phase transition ($a \lesssim 3$), even when including  higher $n$ terms (dotted blue line). In fact, one can show that in order to obtain successful description at the phenomenologically relevant temperatures which are achievable at RHIC and LHC, $1 \lesssim a \lesssim 3$ (yellow shaded band in Fig.~\ref{fig:cs2app}),  one has to include NLO terms in $\Lisn12$ expansion. In Fig.~\ref{fig:cs2app} we present the expansion analogous to (\ref{cs24}) obtained assuming that $\Lisn12 \approx e^{-\s2 a}+e^{-2\s2 a} 2^{-(n+1/2)}$ and taking terms up to  $n=1$ (purple dashed-dotted line)  and  $n=4$ (black dashed-double-dotted line) in (\ref{EneGL}) and (\ref{peGL}). We see that see that taking already first NLO term in $\Lisn12$ is enough for proper description of temperature dependence of the speed of sound close to phase transition. We do not show these, somewhat lengthy, expressions for $c_s^2$ herein. 

\subsection{Shear viscosity}

For the shear viscosity coefficient we use the formula~(\ref{eq:EtaSimplified})
\beq
\eta =\frac{1}{10} \frac{g_0\gamma_\smallG^5}{3\pi^2} \frac{\trel}{T} \int_0^\infty \dd y \frac{\left( y^4-1 \right)^2}{y^4+1} \frac{e^A}{(e^A-1)^2} \,,
\label{eta0}
\eeq
where we have introduced the notation $A = a \sqrt{y^2+y^{-2}}$. This may be further rewritten as
\beq
\eta = \frac{1}{10} \frac{g_0\gamma_\smallG^5}{3\pi^2} \frac{\trel}{T} {\tilde \eta}
\label{eta1}
\eeq
with the dimensionless quantity 
\beq
{\tilde \eta} = \int_0^\infty \dd y \frac{\left( y^4-1 \right)^2}{y^4+1} \frac{e^A}{(e^A-1)^2} \,.
\label{teta0}
\eeq
Changing first the $y$-integration range to the interval \mbox{$1 \leq y \leq \infty$} and, second, changing the integration variable from $y$ to $z$ using  the formulas (\ref{zvar}) we find
\beq
{\tilde \eta}  = \int_1^\infty \dd z f_\eta(z) \frac{e^{\s2 a z}}{(e^{\s2 a z}-1)^2} \,.
\eeq
Here
\beq
f_\eta(z) = \frac{z \, y}{\sqrt{z^4-1}} \frac{(y^4-1)^2 (1+y^{-6})}{y^4+1} \,,
\eeq
with the expansion
\begin{align}
f_\eta(z) &= \sum_{n=1}^\infty \aetan (z-1)^{n-1/2} \nonumber \\
&= 8 \sqrt{z-1} + 34 (z-1)^{3/2} \nonumber\\
&\hspace{1em}+ \frac{111}{4} (z-1)^{5/2} +\frac{257}{16} (z-1)^{7/2} + \cdots \,.
\end{align}
In this way we come to the series of the integrals of the form 
\begin{align}
\label{etabar2}
{\tilde \eta}  &=   \sum_{n=1}^\infty \aetan \int_1^\infty \dd z\, \frac{ (z-1)^{n-1/2}  e^{\s2 a z}}{(e^{\s2 a z}-1)^2} \nonumber \\
&= \sum_{n=1}^\infty \frac{(2n-1) \aetan}{\s2 a} \int_0^\infty \dd \alpha\, \frac{ \alpha^{2(n-1)} }{e^{\s2 a (1+\alpha^2)}-1} \,.
\end{align}
The second line in (\ref{etabar2}) was obtained by the substitution $z=1+\alpha^2$ in the first line and by integration by parts. The integration over $\alpha$ in (\ref{etabar2}) yields the gamma and polylogarithm functions, hence, the final result for the shear viscosity may be written as the following series
\begin{align}
\eta  &= \frac{1}{10} \frac{g_0\gamma_\smallG^5}{3\pi^2} \frac{\trel}{T}  \sum_{n=1}^\infty \aetan \left(\s2 a\right)^{-(n+\frac{1}{2})} \nonumber\\
&\times \Gamma\left(n+\frac{1}{2}\right)  \Lisnm12 \,.
\label{etaGL}
\end{align}
To get (\ref{etaGL}) we used  $(2n-1) \Gamma(n-1/2) = 2 \Gamma(n+1/2)$.

\subsection{Bulk viscosity}

The bulk viscosity coefficient has been given in (\ref{eq:ZetaSimplified}) and can be written as
the sum of two terms
\beq
\zeta = \frac{g_0\gamma_\smallG^5}{3\pi^2} \frac{\trel}{T} \left( c_s^2  \,\, \zeta_1 
-\frac{ \zeta_2}{3} \right).
\eeq
The coefficients $\zeta_i$  ($i=1,2$) are given by the integrals
\beq
\zeta_i  = \int_1^\infty \dd z \, f_{\zeta_i}(z) \frac{e^{\s2 a z}}{(e^{\s2 a z}-1)^2},
\eeq
where the functions $ f_{\zeta_1}$ and $ f_{\zeta_2}$ are defined by the following expressions
\beq
 f_{\zeta_1}(z) &=&  \frac{z}{\sqrt{z^4-1}} (y+y^{-1})  \\
 &=& \sum_{n=0}^\infty \azeta1 (z-1)^{n-1/2} \nonumber \\
&=& \frac{1}{\sqrt{z-1}} + \frac{3 \sqrt{z-1}}{4} -\frac{5}{32} (z-1)^{3/2}  \nonumber \\
&& + \frac{7}{128} (z-1)^{5/2} -\frac{45}{2048} (z-1)^{7/2} +
\cdots , \nonumber
\eeq
\beq
 f_{\zeta_2}(z) &=&  \frac{z\,y}{\sqrt{z^4-1}} \frac{(y^4-1) (1-y^{-2})}{y^4+1}  \\
 &=& \sum_{n=0}^\infty \azetad (z-1)^{n-1/2} \nonumber \\
&=&  2\sqrt{z-1} -\frac{3}{2} (z-1)^{3/2} \nonumber \\
&& +  \frac{23}{16} (z-1)^{5/2} -\frac{91}{64} (z-1)^{7/2} + \nonumber \cdots .
\eeq
Note that the series defining $ f_{\zeta_2}$ starts with the term $2\sqrt{z-1}$, thus the coefficient $a_{\zeta_2}^{(0)}$ vanishes.
Collecting the expressions for $\zeta_1$ and $\zeta_2$ together, we arrive at the formula
\begin{align}
\zeta  &=  \frac{g_0\gamma_\smallG^5}{3\pi^2} \frac{\trel}{T}  
 \sum_{n=0}^\infty \,
\left[ c_s^2(a) \,\azeta1 - \frac{1}{3} \azetad \right] \nonumber\\
&\times \left(\s2 a\right)^{-(n+\frac{1}{2})} \Gamma\left(n+\frac{1}{2}\right)  \Lisnm12.
\label{zetaGL}
\end{align}

\subsection{$\zeta/\eta$ ratio for $T \to 0$}
\label{sec:LowTExpansionZetaEta}

For large $a$ all the polylogarithmic functions can be approximated by their arguments (this corresponds to the limit of Boltzmann statistics). Then, the leading contribution in (\ref{zetaGL}) is obtained by inclusion of the first two terms 
\begin{align}
\zeta  &\approx  \frac{g_0\gamma\smallG^5}{3\pi^2} \frac{\trel}{T}  \frac{e^{-\s2 a}}{\s2 a}
\left[ \left( \frac{1}{\s2 a} \cdot 1 - \frac{1}{3} \cdot 0 \right) \sqrt{\pi}  \right. \nonumber\\
&\hspace{1em}+ \left.\left(  \frac{1}{\s2 a} \cdot \frac{3}{4} - \frac{1}{3} \cdot 2 \right)  \frac{\sqrt{\pi}}{2} \frac{1}{\s2 a}
 \right] \nonumber \\
&\approx   \frac{\s2}{15}\frac{g_0\gamma_\smallG^4\trel}{\pi^{3/2}}   \frac{e^{-\s2 a}}{\left(\s2 a\right)^{1/2}} \, .
\end{align}
Similar expansion for the shear viscosity (with the term $n=1$ alone) gives
\beq
\eta  = \frac{\s2}{9}\frac{g_0\gamma_\smallG^4\trel}{\pi^{3/2}}   \frac{e^{-\s2 a}}{\left(\s2 a\right)^{1/2}} \,,
\eeq
which leads to the ratio
\beq
\frac{\zeta}{\eta} = \frac{5}{3} \,.
\eeq
By comparing this with the scaling relation in Eq.~(\ref{eq:ZetaEtaRatio}) we conclude that $\kappa_\GZ=5$.

Similarly as in Sec.~\ref{sec:epsv} one can go beyond the leading order in polylogarithm expansion to see how the result for $\kappa$ presented above is modified when one includes higher order therms in $a$ expansion. 
\begin{figure}[t]
\centering
\includegraphics[width=0.95\columnwidth]{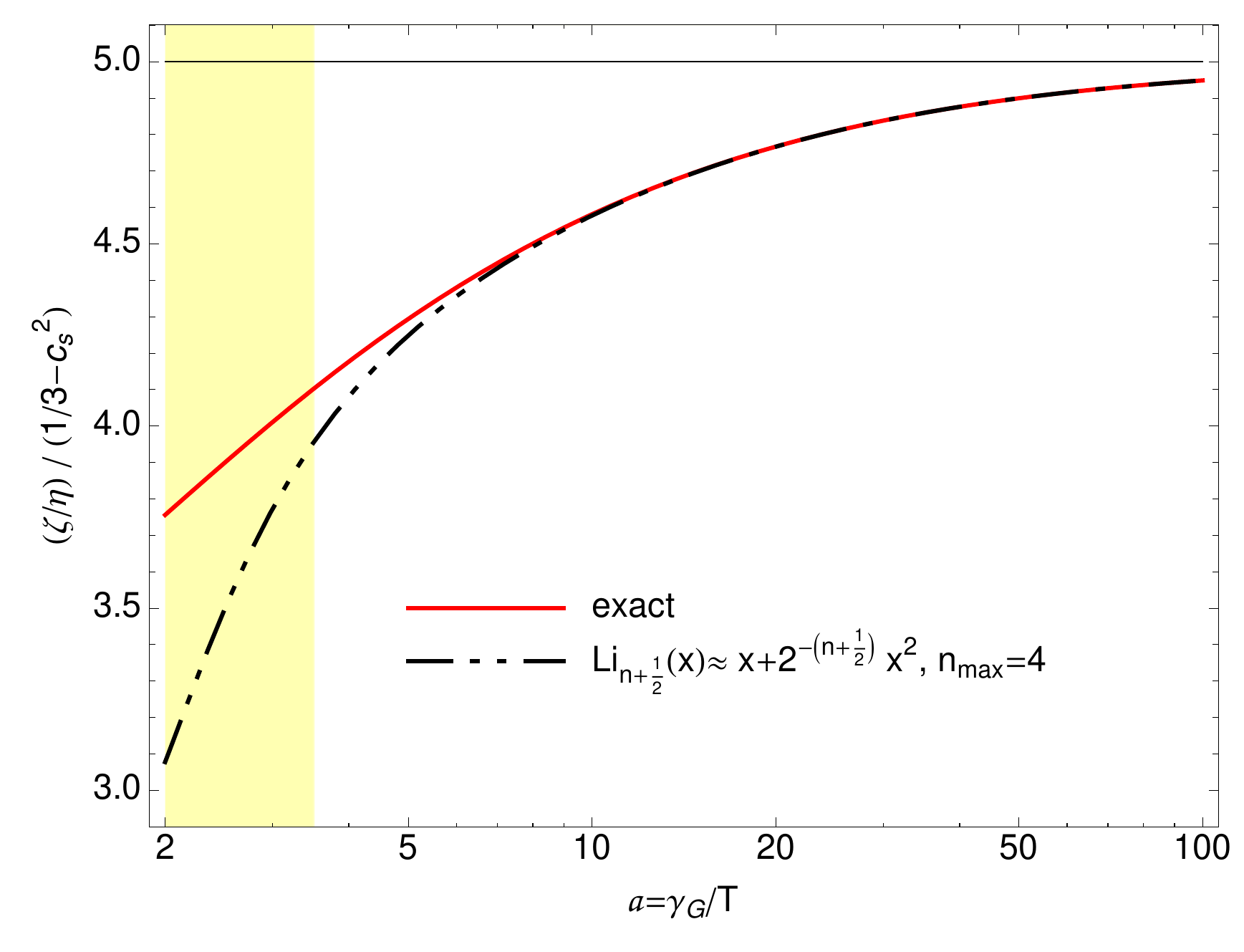}
\caption{\label{fig:kappa} (Color online) The $a=\gamma_\smallG/T$ dependence of  $-\zeta/\eta/(c_s^2-1/3)$ at leading order (solid red line)  and assuming $\Lisn12 \approx e^{-\s2 a}+e^{-2\s2 a} 2^{-(n+1/2)}$ (black dashed-double-dotted line), see discussion of Fig.~\ref{fig:cs2app}.}%
\end{figure}
In Fig.~\ref{fig:kappa} we present the $\kappa=-\zeta/\eta/(c_s^2-1/3)$ at leading order (solid red line) of $\Lisn12$  and assuming $\Lisn12 \approx e^{-\s2 a}+e^{-2\s2 a} 2^{-(n+1/2)}$ (black dashed-double-dotted line). One can observe that close to the phase transition temperature the value of $\kappa$ is significantly smaller than the LO result and somewhat closer to the value obtained in Fig.~\ref{fig:ZetaEtaUniversality}.

\section{Momentum dependent relaxation time}
\label{sec:MomTauRel}

In this Section we give more details describing calculations of the kinetic coefficients of the GZ plasma in the case where the relaxation time depends on the momentum according to Eq.~(\ref{eq:trelAlpha}). To find a high-temperature leading-order term of the bulk viscosity defined by Eq.~(\ref{eq:ZetaSimplifiedAlpha})  we first write
\beq
\zeta = \frac{g_0 \gamma_\smallG^5 \trel^0}{6 \pi^2 T} \,{\tilde \zeta},
\eeq
where
\beq
{\tilde \zeta } =  \int_0^\infty \dd y\,  \left[\frac{1}{3} -c_s^2 -\frac{2}{3 (y^4+1)} \right] \frac{y^\alpha (y^4-1)\, e^{A}}{(e^A-1)^2}.
\nonumber \\ \label{eq:zetabarT}
\eeq
Here we have introduced the notation $A = a \sqrt{y^2 + y^{-2}}$ and $a = \gamma_\smallG/T$. In the high temperature limit, $T~\to~\infty$,  corresponding to $a \to 0$,  we use the approximation 
\beq
\frac{1}{3} -c_s^2 = \frac{\beta_\GZ a^3}{3} , \quad \beta_\GZ = \frac{15}{4 \sqrt{2} \pi^3}.
\label{eq:betaGZ}
\eeq
Substituting Eq.~(\ref{eq:betaGZ}) into Eq.~(\ref{eq:zetabarT}) one finds that the leading term is given by the integral over the variable $y$ which is analytic and yields (for $\alpha \geq 1$)
\beq
{\tilde \zeta } = \frac{\beta}{3 } \, a^{-\alpha-2}\, (4+\alpha) \Gamma(4+\alpha) \zeta(4+\alpha).
\eeq
Here $\Gamma(x)$ is the Euler $\Gamma$ function, while $\zeta(x)$ is the Riemann $\zeta$ function. In an analogous way, in the case of the shear viscosity,  we obtain 
\beq
\eta =  \frac{g_0 \gamma_\smallG^5 \trel^0}{30 \pi^2 T} {\tilde \eta},
\eeq
where the leading term for ${\tilde \eta}$ is (again for $\alpha \geq 1$)
\beq
{\tilde \eta} = a^{-5-\alpha} (4+\alpha) \Gamma(4+\alpha) \zeta(4+\alpha).
\eeq
Collecting the results given above we find the scaling with the first power of the conformal measure given by Eq.~(\ref{eq:scalingAlpha}). We have also found the numerical evidence that the linear scaling holds for $0 < \alpha < 1$.

\section{Massive BE plasma}
\label{sec:MassiveGas}

\subsection{Thermodynamic properties}

In this Section we collect the basic thermodynamic formulas for the massive plasma. Below, the parameter
 $\epsilon = +1$ for bosons and $\epsilon = -1$ for fermions. The quantity $g_0$ is the degeneracy factor connected with internal quantum numbers, and $\mu$ is the chemical potential (set equal to zero in the calculations). The series form follows from the expansion of the BE  or Fermi-Dirac (FD) distributions, $K_n$ are the modified Bessel functions,
\begin{align}
\label{eq:EnDensMassive}
\varepsilon &= \frac{g_0}{2\pi^2}\, T\,m^2\,\epsilon\,\sum_{\kappa=1}^\infty \frac{\epsilon^\kappa}{\kappa^2} e^{\frac{\mu}{T}\kappa} \left[ 3\, T\, K_2\left( \frac{m}{T}\kappa \right)\right. \nonumber\\
&\hspace{1em} + \left. m\, \kappa\, K_1\left( \frac{m}{T}\kappa \right) \right], \\
\label{eq:PressureMassive}
P &=
\frac{g_0}{2\pi^2}\, T^2\,m^2\,\epsilon\, \sum_{\kappa=1}^\infty \frac{\epsilon^\kappa}{\kappa^2} e^{\frac{\mu}{T}\kappa}\, 
K_2\left( \frac{m}{T}\kappa\right), \\
 s &=
\frac{g_0}{2\pi^2}\, m^2\, \epsilon\, \sum_{\kappa=1}^\infty \frac{\epsilon^\kappa}{\kappa^2} e^{\frac{\mu}{T}\kappa} \left[ \left( 4\,T - \mu\,\kappa \right) K_2\left( \frac{m}{T}\kappa \right) \right. \nonumber\\
&\hspace{1em} + \left. m\,\kappa\, K_1\left( \frac{m}{T}\kappa \right) \right] \,.
\end{align}
It is also worth recalling the high temperature behavior of these thermodynamic quantities. At $T \gg m$ we find that
\begin{align}
\varepsilon &= 3 c_\SB T^4 - \frac{2}{3}m^2 T^2  +\mathcal{O}\left(1 \right)\,,\\
P &= c_\SB T^4 - \frac{2}{3}m^2 T^2  +\mathcal{O}\left(1 \right)\,,\\
s &= 4 c_\SB T^3 - \frac{4}{3} m^2 T +\mathcal{O}\left(1 \right) \,,
\end{align}
where $c_\SB = 8\pi^2/45$ for SU(3), which reflect the soft EOS of the massive BE plasma. The first terms in the above expansion are the Stefan-Boltzmann energy density, pressure and entropy, respectively.

\subsection{Kinetic coefficients}

For vanishing chemical potential $\mu = 0$, following the same linearization procedure as above, we obtain the bulk and shear viscosities as
\begin{align}
\label{eq:ZetaMassive}
\zeta &= \frac{g_0 m^5}{6\pi^2} \frac{\trel}{T} \int_0^\infty \dd y \, y^2 \left[ c_s^2-\frac{1}{3}\frac{y^2}{y^2+1} \right]f(1+f) ,\\
\label{eq:EtaMassive}
\eta &= \frac{g_0 m^5}{30\pi^2} \frac{\trel}{T} \int_0^\infty \dd y \, \frac{y^6}{y^2+1} f(1+f) \,,
\end{align}
where $f = [\exp(m \sqrt{y^2+1}/T)-1]^{-1}$, $y=|\k|/m$, and the speed of sound $c_s^2 = \partial P \big/\partial \varepsilon|_{\mu=0}$ is found from Eqs. (\ref{eq:EnDensMassive}) and (\ref{eq:PressureMassive}), respectively. The viscosities, given by Eqs.~(\ref{eq:ZetaMassive}) and (\ref{eq:EtaMassive}), agree with the ones found from \cite{Sasaki:2008fg}.

At high temperatures we can also find the generic behavior of $\zeta/s$ and $\eta/s$. In order to find the leading coefficient of the high-$T$ expansion, it is enough to simply consider the Stefan-Boltzmann entropy $s \approx s_\SB = 4 c_\SB T^3$. Then it is straightforward to find the following expressions at $T \gg m$,
\begin{align}
\frac{\zeta}{s} &= \frac{5}{16 \pi^3} \frac{m^3 \trel}{T^2}  +\mathcal{O}\left(\frac{1}{T^4} \right) \,,\\
\frac{\eta}{s} &= \frac{\trel T}{5} +\mathcal{O}\left(\frac{1}{T} \right) \,,
\end{align}
which results in the ratio
\beq
\frac{\zeta}{\eta} \approx \frac{25}{16\pi^3} \left(\frac{m}{T}\right)^3
\eeq
at high temperatures. This leads directly to Eq.~(\ref{threehalvesscaling}).

\subsection{Momentum dependent relaxation time}
\label{BEmomdep}

We have also studied the BE massive gas with the momentum dependent relaxation time, in the similar way as we did it for the GZ plasma in~Sec.~\ref{sec:MomTauRel}. In this case ($\alpha > 0$)
\beq
\trel = \trel^0 (|\k|/m)^\alpha.
\label{eq:trelBE}
\eeq
Following the same steps as in~Sec.~\ref{sec:MomTauRel} we find
\beq
\zeta = \frac{g_0 m^5 \trel^0}{6 \pi^2 T} \,{\tilde \zeta},
\eeq
where
\beq
\label{eq:zetabarTBE}
{\tilde \zeta } =  \int_0^\infty \dd y\,  \left[\frac{1}{3} -c_s^2 -\frac{1}{3 (y^2+1)} \right] \frac{y^{\alpha+4}\, e^{A}}{(e^A-1)^2}.
\eeq
In this case  $A = a \sqrt{ y^{2}+1}$ and $a = m/T$. In the high temperature limit, $T~\to~\infty$,  we use the approximation
\beq
\frac{1}{3} -c_s^2 = \frac{\beta_\BE a^2}{3} , \quad \beta_\BE = \frac{5}{4 \pi^2}.
\label{eq:betaBE}
\eeq
Substituting Eq.~(\ref{eq:betaBE}) into Eq.~(\ref{eq:zetabarTBE}) one finds 
\beq
{\tilde \zeta } = \frac{a^{-\alpha-3}}{3} \, \, \left[ \beta_\BE H(4+\alpha) - H(2+\alpha) \right], 
\eeq
where $H(x) =  x \Gamma(x) \zeta(x)$. Similarly, in the case of the shear viscosity,  one obtains 
\beq
\eta =  \frac{g_0 m^5 \trel^0}{30 \pi^2 T} {\tilde \eta},
\eeq
where the leading term for ${\tilde \eta}$ is
\beq
{\tilde \eta} = a^{-5-\alpha} H(4+\alpha).
\eeq
Consequently, we obtain the scaling
\beq
\frac{\zeta}{\eta} = 5 \left(1-\frac{H(2+\alpha)}{\beta_\BE H(4+\alpha)} \right) \left(\frac{1}{3}- c_s^2 \right) .
\label{eq:scalingAlphaBE}
\eeq

As the parameter $\alpha$ grows, the ratio $H(2+\alpha)/H(4+\alpha)$ tends to zero, and the result for the BE massive gas agrees with the result for the GZ plasma. On the other hand, for $\alpha=0$ one finds that $H(2)/(\beta_\BE H(4))=1$ and the linear scaling (\ref{eq:scalingAlphaBE}) breaks down. In this case one gets the scaling of the ratio of the viscosities with the power $3/2$, in agreement with our result given by Eq.~(\ref{threehalvesscaling}).

Finally, we note that we have also done the calculations for the massive Boltzmann gas. Interestingly, for $\alpha=1$ we reproduce a linear scaling with the prefactor 2, which agrees with the recent result obtained in Ref.~\cite{Jaiswal:2016sfw}.

\bibliography{GZtransport}
\end{document}